\documentclass[]{aastex62}

\shorttitle{Radial velocity photon noise for F--M dwarfs}
\shortauthors{Reiners \& Zechmeister}

\begin{document}

\title{Radial velocity photon limits for the dwarf stars of spectral
  classes F--M}

\correspondingauthor{Ansgar Reiners}
\email{Ansgar.Reiners@phys.uni-goettingen.de}

\author{Ansgar Reiners}
\affil{Institut f\"ur Astrophysik \\
Friedrich-Hund Platz 1\\
D-37077 G\"ottingen, Germany}

\author{Mathias Zechmeister}
\affil{Institut f\"ur Astrophysik \\
Friedrich-Hund Platz 1\\
D-37077 G\"ottingen, Germany}



\begin{abstract}

  The determination of extrasolar planet masses with the radial
  velocity (RV) technique requires spectroscopic Doppler information
  from the planet's host star, which varies with stellar brightness
  and temperature. We analyze Doppler information in spectra of F--M
  dwarfs utilizing empirical information from HARPS and CARMENES, and
  from model spectra. We come to the conclusions that an optical setup
  ($BVR$-bands) is more efficient that a near-infrared one ($YJHK$) in
  dwarf stars hotter than 3200\,K.

  We publish a catalogue of 46,480 well-studied F--M dwarfs in the
  solar neighborhood and compare their distribution to more than one
  million stars from Gaia DR2. For all stars, we estimate the RV
  photon noise achievable in typical observations assuming no activity
  jitter and slow rotation. We find that with an ESPRESSO-like
  instrument at an 8m-telescope, a photon noise limit of
  10\,cm\,s$^{-1}$ or lower can be reached in more than 280 stars in a
  5\,min observation. At 4m-telescopes, a photon noise limit of
  1\,m\,s$^{-1}$ can be reached in a 10\,min exposure in approx.\
  10,000 predominantly sun-like stars with a HARPS-like (optical)
  instrument. The same applies to $\sim$3000 stars for a red-optical
  setup covering the $RIz$-bands, and to $\sim$700 stars for a
  near-infrared instrument. For the latter two, many of the targets
  are nearby M dwarfs. Finally, we identify targets in which
  Earth-mass planets within the liquid water habitable zone can cause
  RV amplitudes comparable to the RV photon noise. Assuming the same
  exposure times, we find that an ESPRESSO-like instrument can reach
  this limit for 1\,M$_\Earth$ planets in more than 1000 stars. The
  optical, red-optical, and near-infrared configurations reach the
  limit for 2\,M$_\Earth$ planets in approximately 500, 700, and 200
  stars, respectively.

\end{abstract}

\keywords{catalogs --- instrumentation: spectrographs --- planets and satellites: detection --- stars: low-mass --- stars: solar-type --- techniques: radial velocities}


\section{Introduction} \label{sec:intro}

The search for planets around other stars, the characterization of
other planetary systems, and the quest for other planets similar to
Earth are motivation for a broad range of astronomical
research. During the last decades, astronomical techniques improved so
that today we know about the existence of thousands of planets, and
for many we can determine their masses, sizes, and potential
atmospheric composition. A number of complementary techniques exist to
obtain information about planets and their characteristics
\citep[e.g.,][]{2014exha.book.....P}. Most successful (in number of
planet discoveries) are the transit technique that, among other
parameters, measures a planet's size, and the radial velocity (RV)
technique that determines the planet's (projected) mass. Ideal targets
are those that can be measured with both the RV and the transit
technique. A large number of systems applicable to both methods is
expected from the satellite transit missions TESS
\citep{2015JATIS...1a4003R} and PLATO \citep{2014ExA....38..249R}.

The mass of a planet is determined via reflex motion of the star
caused by an orbiting planet. Typical velocities of the star's motion
are several ten m\,s$^{-1}$ for giant planets (12\,m\,s$^{-1}$ for
Jupiter in the solar system), and several cm\,s$^{-1}$ for terrestrial
planets with masses similar to Earth (9\,cm\,s$^{-1}$ for Earth around
the Sun). Current instrumental limitations together with intrinsic
stellar variability limits the RV technique to planets that cause
stellar velocity amplitudes on the order of 1\,m\,s$^{-1}$ or larger
\citep{2016PASP..128f6001F}. Instruments like ESPRESSO
\citep{2014AN....335....8P, 2019arXiv190804627H} aim for the
10\,cm\,s$^{-1}$-limit, and a new generation of spectrographs at
telescopes beyond the 10m-class are planned with similar perfomance
goals \citep{2016SPIE.9908E..23M, 2016SPIE.9908E..22S}.

With the RV method, planets are easier to detect if they orbit less
massive stars. Lower-mass dwarf stars are smaller than more massive
stars, and they have cooler surface temperatures. Thus, they are
significantly fainter than warmer stars, and they show a very
different spectral energy distribution exhibiting more flux at longer
wavelengths. They generally show more spectral features because of
molecular absorption. The question about the minimum detectable planet
mass around any given star is therefore not easily answered. This is
the topic of this paper.

The RV precision achievable in a stellar spectrum depends on the
number of photons (or the signal-to-noise ratio, SNR) and the amount
of spectroscopic information, i.e., the presence of spectral
features. The RV information content of stellar spectra at visible
wavelengths was discussed for example by \citet{1985Ap&SS.110..211C,
  1996PASP..108..500B}, and \citet{2001A&A...374..733B}. They
demonstrated that sun-like stars are ideal targets for RV
determination at optical wavelengths. After the realization that the
possibility of life on planets around M dwarfs cannot be generally
ruled out \citep{2007AsBio...7...30T, 2007AsBio...7...85S}, they
became a second focus of RV surveys because smaller planets can be
discovered around them. Also, low-mass dwarfs are the most numerous
stars in our galactic neighborhood \citep{2006AJ....132.2360H,
  2016AAS...22714201H}. Therefore, growing effort is spent to search
for our closest neighbor planets with several RV instruments
\citep{2017RNAAS...1a..51W}. Additional motivation comes from the
expectation of transiting planets discovered with TESS, and it is
important to identify the optimal strategy for RV survey and follow-up
observations.  \citet{2018AJ....156...82C} presented a specialized
calculation of the total observing time required to measure planet
masses from the expected TESS planet yield. In addition to photon
noise, they took into account other mechanism causing RV noise
including stellar activity. This so-called RV jitter acts as an
additional source of noise that can be estimated from Ca activity
\citep[e.g.,][]{2005PASP..117..657W, 2000A&A...361..265S}, photometric
flicker \citep{2014ApJ...780..104C}, or measurements of stellar
rotation \citep{2017A&A...600A..13A}. The amplitude of RV jitter is
expected to depend on wavelength, which can be useful to
distinguishing between Keplerian motion and the stellar activity in RV
measurements \citep{2018A&A...609A..12Z}. However, the wavelength
dependence of stellar activity on RV measurements is not well
understood. Temperature spots are expected to cause RV jitter that is
smaller at longer wavelengths because of diminished contrast between
hot and cool regions \citep[e.g.,][]{2006ApJ...644L..75M,
  2008A&A...489L...9H, 2010ApJ...710..432R}. Zeeman broadening, on the
other hand, could lead to an increase of RV jitter with wavelength
\citep{2013A&A...552A.103R}. \citet{2018AJ....156...82C} therefore
chose to implement a wavelength independent term for RV jitter. They
highlight the comparable performance of optical and near-IR
spectrographs albeit they are not considering spectrographs that cover
the red optical wavelength range between 700 and 900\,nm.

Identifying the best strategy for RV measurements in low-mass stars
was hampered by the difficulty of synthesizing high-resolution spectra
including all relevant molecular features, and by the lack of
high-resolution infrared observations. Investigation of RV information
carried out by, e.g., \citet{2010ApJ...710..432R, 2011A&A...532A..31R,
  2013PASP..125..240B, 2015arXiv150301770P, 2015PASP..127.1240B,
  2016A&A...586A.101F}, showed that the RV observations in M dwarfs
should focus on wavelengths redder than 600--700\,nm but their results
depended on the models' assumptions. \citet{2018AJ....155..198A}
employed observations of Barnard's star (M4V) obtained with HARPS,
ESPADONS, and CRIRES to test the radial velocity content of M dwarf
spectra. \citet{2018A&A...612A..49R} determined the amount of RV
information in M dwarf spectra between the $V$- and $H$-bands from
observations of 324 stars obtained with the CARMENES instrument
\citep{2016SPIE.9908E..12Q}. In this paper, we revisit the radial
velocity content of M dwarf spectra, and we add empirical information
about M dwarf and hotter star spectra at optical wavelengths. As our
main product, we compute the RV precision limit set by photon noise
that can be reached with any (ideal) spectrograph and target
brightness for an extensive sample of F--M dwarf stars.

We define a sample of stars with well characterized properties in
Section\,\ref{sect:sample}, and we discuss in
Section\,\ref{sect:detprec} how the values of RV precision are
calculated. In Section\,\ref{sect:catalogue}, we introduce our RV
precision catalogue, and we discuss the performance and the potential
of different instrument designs regarding the discovery of low-mass
planets. In Section\,\ref{sect:hablim}, we look into the detectability
of planets inside the liquid water habitable zone. As a tool for
optimizing instrument design, and planning RV observations, we provide
an online-calculator for the RV precision achievable for a given star
and instrument setup (see Section\,\ref{sect:detprec}).

\section{The stellar sample}
\label{sect:sample}

In our sample, we include catalogues with main sequence stars of
spectral types F--M. We focus on catalogues with precise information
about surface temperature and stellar mass. We summarize the
catalogues included in our work in Table\,\ref{tab:catalogues} where
we provide the numbers of stars as well as temperature and brightness
ranges. For stars contained in more than one catalogue, we give
preference to the publications in the order as listed in
Table\,\ref{tab:catalogues}. For F-, G-, and K-dwarfs, we include
\emph{The Catalog of Earth-Like Exoplanet Survey Targets}
\citep{2016AJ....151...59C}, \emph{The Geneva-Copenhagen survey of the
  Solar neighbourhood} \citep{2004A&A...418..989N}, and the catalogue
on \emph{Spectral Properties of Cool Stars}
\citep{2016ApJS..225...32B}. For cooler stars, data are taken from
\emph{The catalogue of nearby cool host-stars}
\citep{2014MNRAS.443.2561G}, \emph{The SPIRou Input Catalogue}
\citep{2018MNRAS.475.1960F}, and \emph{The Near-Infrared Spectroscopic
  Survey of 886 Nearby M Dwarfs} \citep{2015ApJS..220...16T}. These
catalogues all provide spectroscopic determinations of stellar
temperature together with magnitudes in different
filters. \citet{2018MNRAS.475.1960F} provide $H$-magnitudes but no
$J$-magnitudes and we approximate $J = H$. We neglect five of the 2974
stars from \citet{2014MNRAS.443.2561G}, because they are reported to
have extremely low temperatures that appear to be outside the range
for which the derived parameters from that work are valid.  We
complement this information with a sample of very late-M dwarfs by
\citet{2009ApJ...705.1416R}, for which we estimate $T_{\rm eff}$ from
spectral type according to the relations provided in
\citet{1995ApJS..101..117K} and \citet{2004AJ....127.3516G}.  Overall,
our catalogue contains 46,480 stars from these detailed
catalogues. Targets of the CARMENES input catalogue
\citep{2018A&A...612A..49R} are almost all contained in the list of
catalogues above and therefore not explictly added.

\begin{deluxetable*}{lcccc}[h]
  \tablecaption{Stellar catalogues used for this study prioritized
    from top to bottom. The second column provides the number of stars
    adopted from each catalogue. Numbers in brackets show the total
    numbers of stars given in each catalogue.\label{tab:catalogues}}
  \tablecolumns{5} 
  \tablewidth{0pt} 
  \tablehead{ \colhead{Catalog} & \colhead{\# stars} &
    \colhead{$T_{\rm eff}$ range} & \colhead{$V$ range} & \colhead{$J$
      range}}
  \startdata 
         Chandler et al. (2016)  &      37354 & 3042 \dots 7199 &  2.4 \dots 13.7 \\
    Nordstr\"om et al. (2004)  &  4782 (14695) & 4613 \dots 7396 &  0.4 \dots 12.7 \\
          Brewer et al. (2016) &   500 (  971) & 4702 \dots 6674 &  1.7 \dots 10.0 \\
          Gaidos et al. (2014) &       2969 & 2700 \dots 4803 &  6.7 \dots 17.8 &  3.9 \dots  9.0 \\
        Fouqu\'e et al. (2018) &  172 (447) & 2656 \dots 4718 &  6.9 \dots 16.6 &  3.3 \dots 12.3 \\
      Terrien et al. (2015)    &  642 (886) & 3276 \dots 4523 &  6.8 \dots 15.9 &  3.9 \dots 12.4 \\
       Reiners \& Basri (2009) &         63 & 2350 \dots 2620 &   &  8.9 \dots 13.3 \\
                     Gaia DR 2 &    1188603 & 3344 \dots 8000 &  3.0 \dots 12.0 \\

  \enddata
\end{deluxetable*}

We use distances from Gaia DR2 \citep{2016A&A...595A...1G,
  2018arXiv180409365G} for all stars for which Gaia DR2 reports
significant parallax measurements (2$\sigma$). Furthermore, we
selected more than 1\,Million targets from Gaia DR2 data using
temperature estimates from \citet{2018arXiv180409374A} with apparent
Gaia-magnitudes brighter than $g = 12$\,mag. For our magnitude plots,
we approximate $g = V$. Although these temperatures are of much lower
precision than the spectroscopically determined ones, we include them
in our analysis to provide an overview about the distribution of stars
that is independent of the selected catalogues.\footnote{We do not
  provide values of RV photon noise for the Gaia DR2 targets in
  Table\,\ref{tab:master}, because they have rather large
  uncertainties.}

\begin{figure}
  \plotone{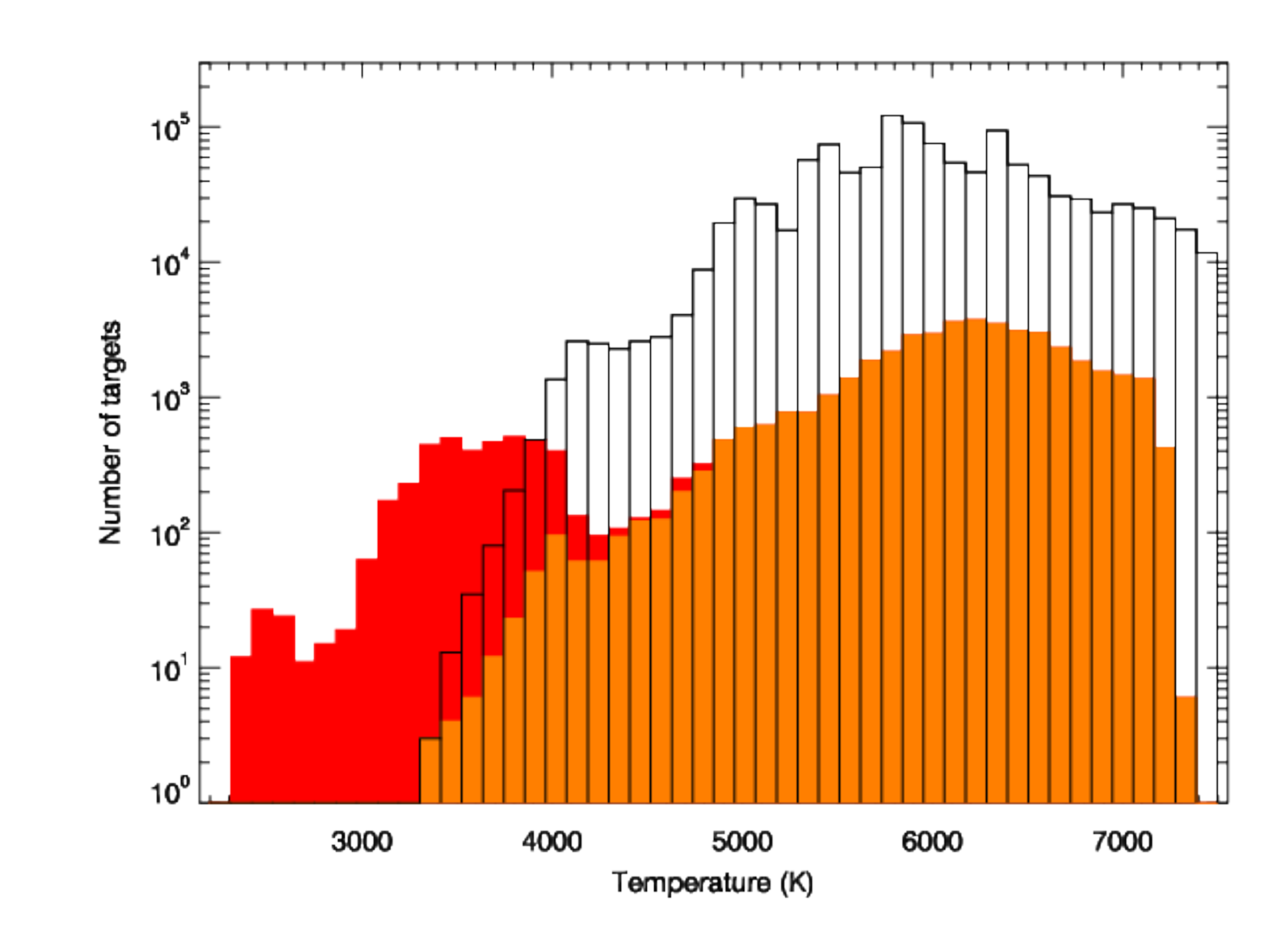}
  \caption{\label{fig:Star_Histo}Histogram of the stars contained in
    the spectroscopic catalogues from
    Table\,\ref{tab:catalogues}. Stars from F-, G-, and K-star
    catalogues are shown in orange \citep{2016AJ....151...59C,
      2004A&A...418..989N, 2016ApJS..225...32B}. Stars from the cool
    dwarf catalogues are added cumulatively in red
    \citep{2014MNRAS.443.2561G, 2018MNRAS.475.1960F,
      2015ApJS..220...16T, 2009ApJ...705.1416R}.  The Gaia DR2 sample
    is overplotted with no color filling.}
\end{figure}

We show the distribution of stars from these catalogues as a function
of temperature in Fig.\,\ref{fig:Star_Histo}. Except for Gaia DR2, the
catalogues reveal an obvious lack of K-dwarfs with temperatures
between 4000\,K and 5000\,K because most of the detailed spectroscopic
work has been done for either M- or G-dwarfs. The distribution of
stars in Gaia DR2 shows that this is a systematic feature of the F--K
catalogues; most of the volume-limited RV survey samples include only
very few K-dwarfs.

For the calculation of planetary mass detection limits, we also need
to estimate the mass of the star. For most of the stars, mass
estimates are provided in the catalogues. For the stars from
\citet{2018MNRAS.475.1960F} and \citet{2009ApJ...705.1416R}, we
estimate mass from temperature according to the 2\,Gyr evolutionary
track from \citet{1998A&A...337..403B}. We did not estimate masses for
the Gaia DR2 data because the uncertainties would be too high for a
meaningful investigation. An additional factor for the estimate of RV
precision is projected rotation velocity $v\sin{i}$.\footnote{For
  resolved spectral lines and rotationally dominated line broadening,
  RV photon noise scales proportional to $(v\,\sin{i})^{1.5}$. In
  practice, the scaling is less steep for most sun-like and cooler
  stars with $v\,\sin{i} < 10$\,km\,s$^{-1}$ and depends on spectral
  resolution \citep[see Fig.\,7 in][]{2010ApJ...710..432R}. } For our
calculations, we do not include the effect of rotational broadening
because its consequences are relatively small for the majority of
stars that are slow rotators, and information on $v\sin{i}$ are
missing for large parts of our targets. The assumption of slow
rotation is not valid for mid- and late-M dwarfs
\citep[e.g.,][]{2018A&A...614A..76J}. Thus, our estimates of RV photon
noise are lower bounds for the RV precision that can be achieved. For
the calculation of the habitable zone distance, we also need the
luminosity of the star. If luminosity is not provided in the
catalogue, we estimate its value from temperature and radius. In
catalogues where the radius is not provided either, we estimate radius
from stellar mass assuming $M/M_{\odot} = R/R_{\odot}$, which is a
reasonable approximation for dwarf stars \citep{2009A&A...505..205D}.

\begin{figure*}
  \plottwo{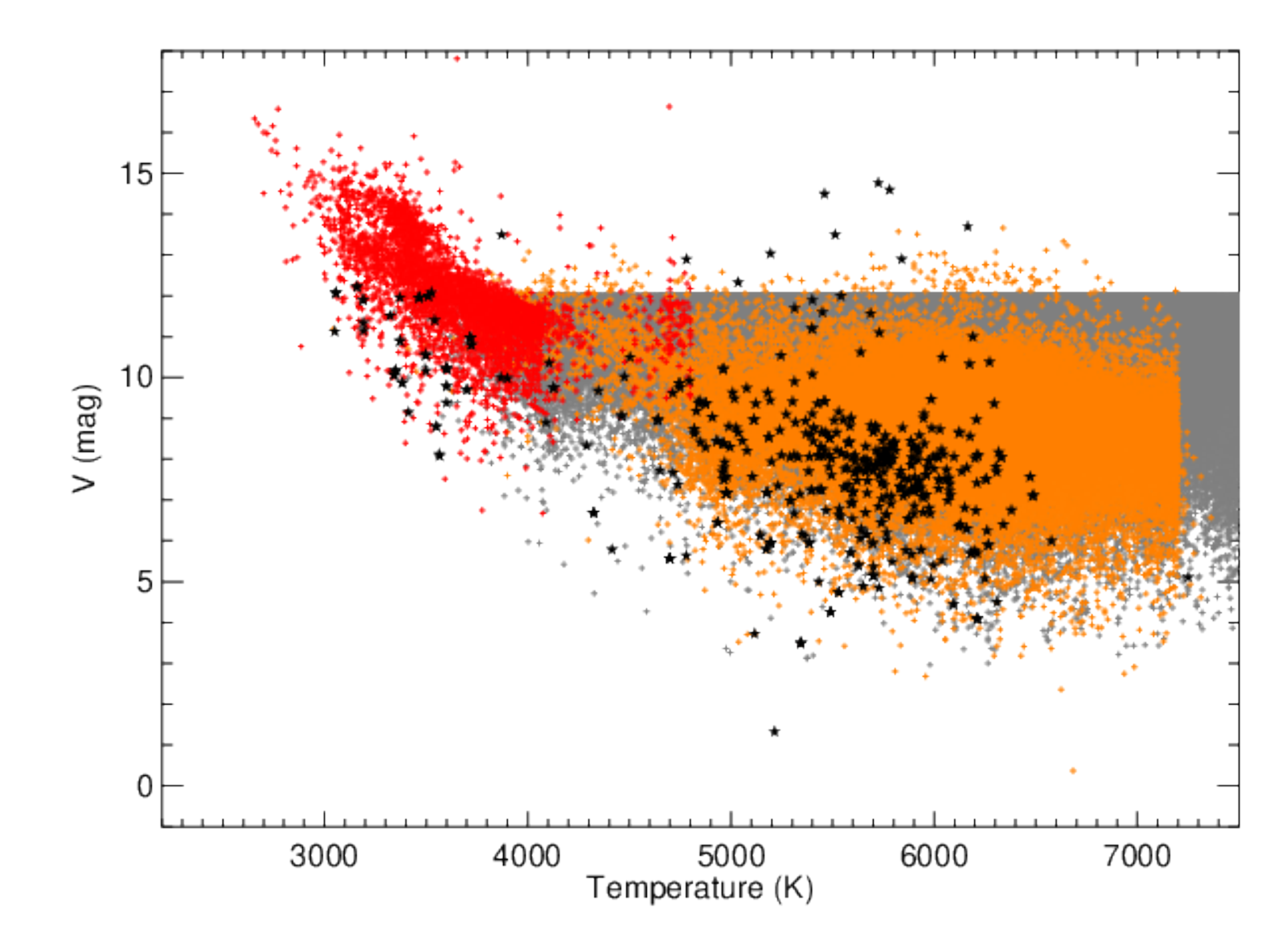}{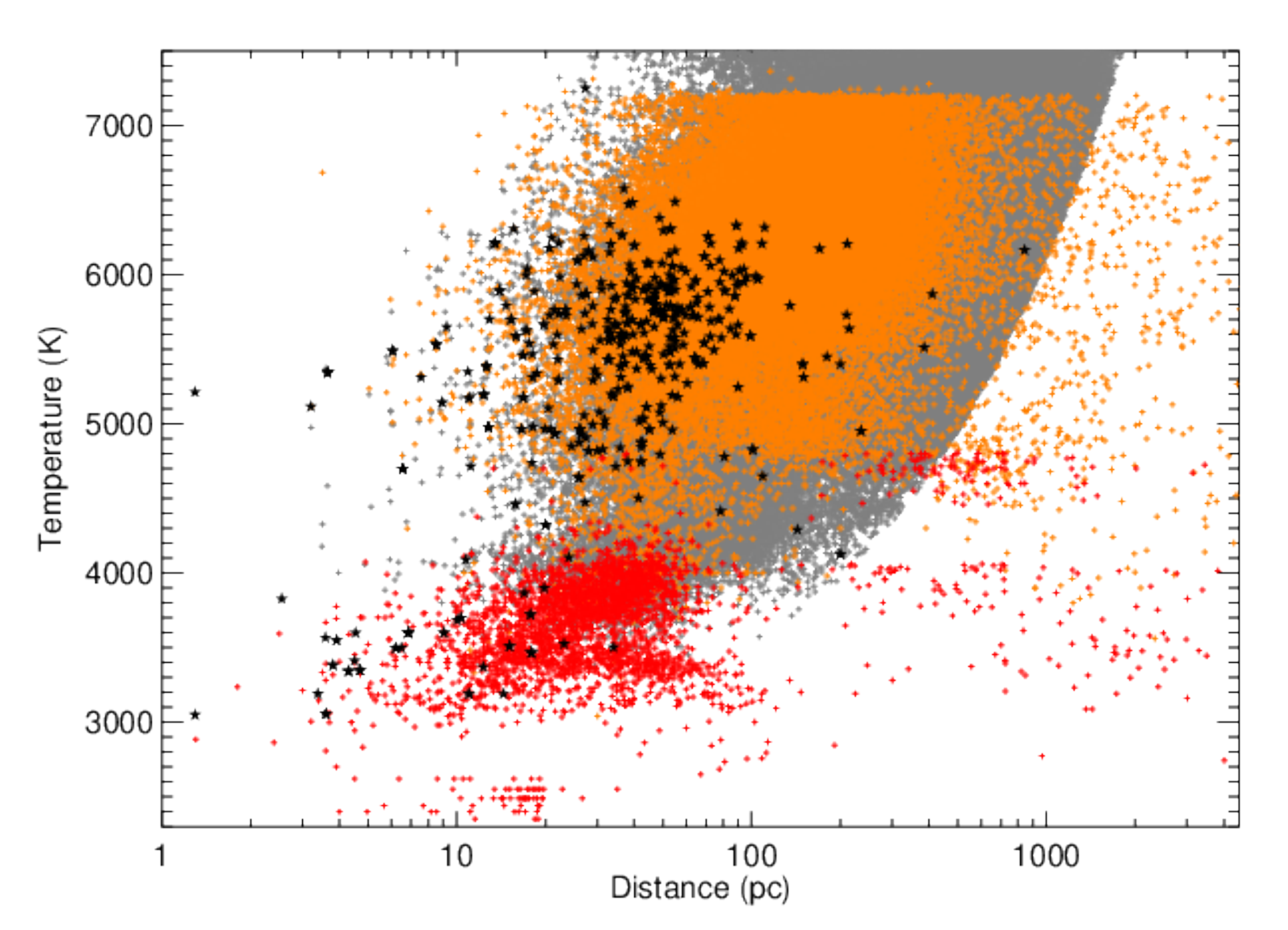}
  \caption{\label{fig:HRD}Sample of stars used for this work with the
    colors for sun-like star catalogues in orange and the cool dwarf
    catalogues in red, respectively, as in
    Fig.\,\ref{fig:Star_Histo}. Stars from Gaia DR2 and
    \citet{2018arXiv180409374A} are shown in grey. Black stars show
    known planet hosts from exoplanet.eu. \emph{Left:} Apparent
    brightness as a function of spectroscopic
    temperature. \emph{Right:} Temperature as a function of distance.}
\end{figure*}

We show the $V$-magnitude of all stars in our sample as a function of
temperature in the left panel of Fig.\,\ref{fig:HRD}, and we plot
temperature as a function of distance in the right panel of that
figure. The two figures provide an orientation about the amount of
stars in the solar neighborhood and about their apparent
brightness. For comparison, we include information on stars for which
extrasolar planets are reported on \emph{The Exoplanet
  Encyclopedia}\footnote{\url{http://exoplanet.eu}} using the stellar
parameters provided there. For these stars, we chose to plot both
values from our spectroscopic catalogues (Table\,\ref{tab:catalogues})
and those from exoplanet.eu. Thus, the stars from exoplanet.eu may be
plotted twice but the parameters given in the literature are not
necessarily identical.

\section{Determining radial velocity photon noise}
\label{sect:detprec}

The fundamental limit for the RV precision that can be achieved from
an observation of any given star is the RV photon noise. In order to
estimate the RV photon noise, we need to specify the RV information
content for the spectrum of a star of given temperature, and its
signal-to-noise ratio, SNR \citep{1985Ap&SS.110..211C,
  1996PASP..108..500B, 2001A&A...374..733B}. The RV precision achieved
in an actual observation can further be deteriorated by additional
line broadening, in particular by stellar rotation and activity. Line
broadening diminishes the amount of spectroscopic features and its
effect can be compensated with higher SNR (longer exposures). Stellar
activity can be a source of additional (additive) noise that cannot be
overcome by longer exposures alone but must be addressed by modeling
the effect using additional information
\citep[e.g.,][]{2012MNRAS.419.3147A, 2016MNRAS.457.3637H,
  2018A&A...620A..47D, 2019MNRAS.487.1082C}. Stellar metallicity
influences the amount of spectral features and also the colors of
stars and therefore the distribution of SNR across wavelengths. In our
estimates, we neglect the influence of metallicity. For the model
calculations, including the effect of metallicity would be relatively
straightforward. Comparison to observed spectra, however, is hampered
by uncertainties of the observed stars' metallicity, particularly in
low-mass stars where the model spectra are known to be imprecise
\citep[see][]{2018AJ....155..198A, 2018A&A...615A...6P}. Relative to
the other effects considered, metallicity is expected to create only
minor deviations for the stars considered in this paper.

Our goal is to calculate RV photon noise for observations taken with
visual light and infrared spectrographs. For this, we investigate the
RV information at wavelengths $\lambda$\,=\,380--2380\,nm. We limit
our study to stars between temperatures $T_{\rm eff} = 2800$\,K and
7000\,K. While RV information can be directly computed from synthetic
models, \citet{2018AJ....155..198A} and \citet{2018A&A...612A..49R}
showed that model spectra do not always correctly represent the
abundance and depth of absorption features. This is particularly
important for absorption from molecular bands in cool
stars. \citet{2018AJ....155..198A} investigated HARPS, ESPADONS, and
CRIRES spectra of Barnard's star (M4). \citet{2018A&A...612A..49R}
employed data from the CARMENES survey to determine RV photon noise
across the M dwarf spectral range (M0--M9). In general, the results of
both studies agree very well; they both conclude that the wavelength
range around 700--900\,nm ($I$-band) provides the most accurate
measurements. However, there are some significant differences at other
wavelength bands that we discuss in the following.

In order to construct a consistent set of empirical photon noise limits across
a variety of stellar temperatures and wavelengths, we make use of several
thousand spectra observed with the HARPS and CARMENES instruments. For M-stars
and wavelengths 550--1780\,nm, we use the RV photon noise values from
\citet{2018A&A...612A..49R} from the CARMENES survey for planets around M
dwarfs. To this data, a telluric mask excluding atmospheric features deeper
than 5\,\% was applied \cite[see][]{2018A&A...609A..12Z}. For hotter stars and
wavelengths 380--650\,nm, we calculate the RV photon noise from ESO-HARPS
archival data from 383 stars. For each star, we take the available archive
data from
HARPS-DRS\footnote{\url{http://www.eso.org/sci/facilities/lasilla/instruments/harps.html}}
including SNR. From this data, we determine the amount of RV information for
each individual order and calculate values of RV precision for stars with
temperatures between $T_{\rm eff} = 3200$\,K and 6000\,K. For all stars and
wavelength ranges in our grid, we additionally compute RV information from
synthetic PHOENIX model spectra \citep{2013A&A...553A...6H} adapted to the
instrument resolutions. For this, we choose to exclude pixels with more than
10\,\% telluric absorption.  We emphasize that there is a conceptual
difference between the RV photon noise limits estimated from theoretical as
well as observed spectra \citep[as done for HARPS and model spectra in][and
here for model spectra only]{2018AJ....155..198A}, and the empirically
determined RV precision \citep[calculated for CARMENES data in][and here for
HARPS data]{2018A&A...612A..49R}. For the empirical case, we use the actual
uncertainties derived from the determination of RVs. We refer to
\citet{2018A&A...609A..12Z} and \citet{2018A&A...612A..49R} for a detailed
discussion. Because of major uncertainties, e.g., in the scaling of SNR (or
the brightness of stars as a function of wavelength), the treatment of
telluric lines, in metallicities, and in the synthetic models, we expect some
disagreement between the RV photon noise limits from different sources. Our
comparison should help to identify these uncertainties and potential problems
in RV precision estimates.

\begin{figure*}
\plotone{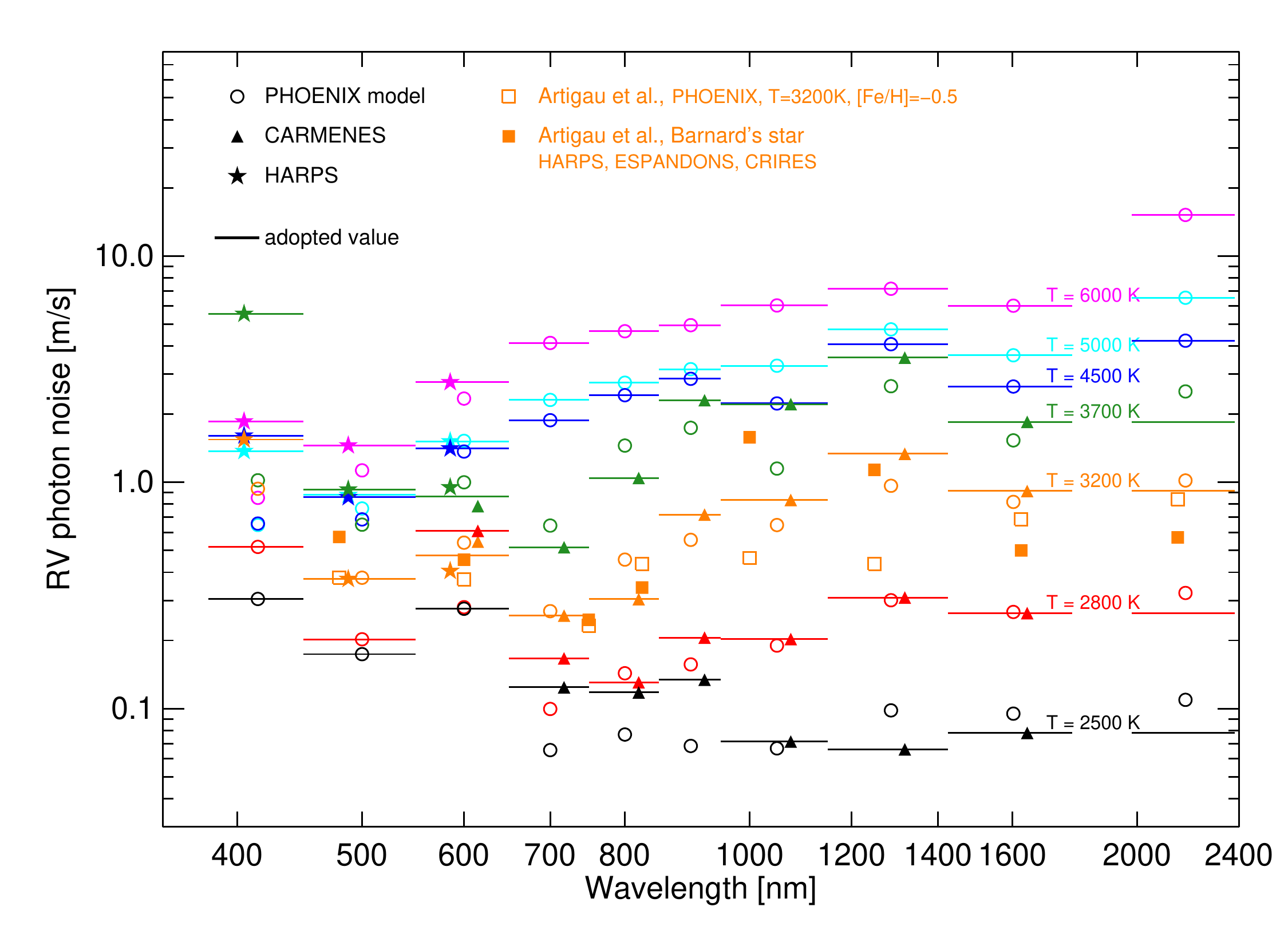}
\caption{\label{fig:rvlambda}Radial velocity photon noise as a
  function of wavelength for spectra from stars of different
  temperatures observed with SNR~=~100 in the $V$-band.  We show RV
  photon noise from difference sources: predicted values from spectral
  models (open circles), CARMENES empirical values (triangles), HARPS
  empirical values (stars), and model predictions (open squares) as
  well as empirically determined values (solid squares) for Barnard's
  star from \citet{2018AJ....155..198A}. Values adopted for our RV
  precision calculations are shown as horizontal lines that also
  indicate our wavelength chunks (Table\,\ref{tab:precision}). Not
  included are results for $T = 7000$\,K (see text).}
\end{figure*}

The results about our RV photon noise limits and empirical precision
estimates for different stellar types are shown in
Fig.\,\ref{fig:rvlambda}. We show our results from CARMENES, HARPS,
and model spectra with different symbols for comparison, and we
include results from \citet{2018AJ....155..198A} for Barnard's
star. The plotted values of RV photon noise are assuming a SNR of 100
per pixel in the $V$-band for $R = 100,000$.\footnote{For the model
  spectra, we define the SNR of a wavelength range as the
  90th-percentile of the SNR for all pixels within that range.}  For
the synthetic model calculations, we use a 2.5-pixel sampling. Note
that this assumption leads to extremely low photon noise values (very
high precision) for very cool (M-type) stars. Observations of M stars
typically do not achieve such high $V$-band SNRs, and
\citet{2018A&A...612A..49R} provide RV photon noise for SNR~=~150 in
the $J$-band. We scale the CARMENES values assuming that the $J$-band
SNR for an object of a given $J$-magnitude is 72\,\% of the $V$-band
SNR from an object that has the same magnitude in the
$V$-band.\footnote{The square-root of the ratio between the number of
  photons from a 10,000\,K blackbody at 550\,nm and 1215\,nm, assuming
  constant instrument resolution, is 0.72. The number of photons per
  resolution element is proportional to the Planck function in units
  per wavelength $(\lambda^{-1})$ times $\lambda^2$.}  From
\citet{2018AJ....155..198A}, we adopt their results for the 2\,\%
telluric absorption limit and scale the numbers from their Table\,4 by
a factor of 7.5, which is approximately the ratio between the SNR of a
M4 star in the $V$-band and in the $J$-band
\citep[see][Fig.\,7]{2018A&A...612A..49R}. To account for rotational
broadening in the synthetic model spectra, we assign rotation rates to
our model stars, which are the lowest ones typical for a given stellar
mass (thus allowing the highest possible RV precision). We estimate
the upper envelope of rotation periods as a function of stellar mass
from Eq.\,(6) in \citet{2017ApJ...834...85N} and Fig.\,8 in
\citet{2013A&A...560A...4R}, and we broaden our synthetic spectra
before calculating RV photon noise. Inactive dwarf stars cooler than
the Sun typically rotate at rotation rates of several 10\,d or slower,
which is less than 1--2\,km\,s$^{-1}$ and negligible for our
calculations. More massive stars, however, rotate substantially
faster, which leads to a significant decrease in RV precision in these
stars. The RV photon noise for the case $T = 7000$\,K is not shown in
Fig.\,\ref{fig:rvlambda}. The approximate lower limit of rotational
velocities for these stars is $v_{\rm eq} = 70$\,km\,s$^{-1}$, and the
typical RV photon noise limit is several 10\,m\,s$^{-1}$
(Table\,\ref{tab:precision}). We note that stars observed under low
inclination angles may show lower projected velocities, $v\,\sin{i}$,
and therefore show lower RV photon noise.

Our first observation in Fig.\,\ref{fig:rvlambda} is that information
on RV photon limits and empirical values on RV precision from
different sources are consistent for most of our parameter space. The
high overall agreement of the \emph{absolute} values from different
methods demonstrates that understanding of RV photon noise has become
relatively robust. In particular, values where CARMENES and HARPS
results are available, i.e., at $\lambda = 600$\,nm and $T_{\rm eff} =
3200, 3700$\,K, the values agree very well. This is an important
validation of our method combining CARMENES and HARPS
measurements. The results from spectral models also show remarkably
similar trends in wavelength and temperature; differences from the
measured values are smaller than a factor two with very few
exceptions. The main reason for discrepancies in the coolest stars is
probably incomplete understanding of line formation, in particular for
molecular lines in M dwarfs \citep[see][]{2018A&A...612A..49R,
  2018AJ....155..198A}, and treatment of telluric lines.

The comparison between the results for Barnard's star from
\citet{2018AJ....155..198A} and our empirical case for 3200\,K also
shows very good agreement in general. At wavelengths between 550\,nm
and 900\,nm, and in the $J$-band, the values only differ by a few
percent. In the $Y$-band around 1\,$\mu$m, our CARMENES data indicate
significantly lower RV photon noise than the CRIRES results, and in
the $H$-band, the CRIRES results suggest lower RV photon noise than
the CARMENES results. As discussed in \citet{2018AJ....155..198A},
this has implications for the design of infrared spectrographs, in
particular whether RV infrared surveys should concentrate on the
$Y$-band or the $H$- and $K$-bands. We cannot provide a clear answer
as to why our results are different from theirs at infrared
wavelengths. The difference in the $H$-band can perhaps be explained
by differences in SNR scaling and in the exact definition of the
wavelength band used for the computations; Fig.\,2 in
\citet{2018AJ....155..198A} shows that most of the RV information in
the $H$- and the $K$-bands is located at the extreme ends of the two
bands. For the $Y$-band, \citet{2018AJ....155..198A} argue that the
models overpredict the achievable RV precision in Barnard's star by a
factor of roughly three, which we cannot confirm. A factor of three in
RV photon noise corresponds to almost an order of magnitude difference
in spectral information, which means a model overprediction of line
depth or density. The spectral atlas of Luyten's star (M3.5) displayed
in Appendix~A of \citet{2018A&A...612A..49R} shows that the observed
lines are in fact generally weaker than predicted in this
star. However, in the $Y$-band, most of the molecular absorption
appears at the predicted location indicating that a correction by a
factor of three is unlikely. Interestingly, the comparison between our
model spectra noise limits and those from \citet{2018AJ....155..198A}
also shows some significant differences, despite both studies use
identical models. We cannot explain this discrepancy but we suspect
that the choice of wavelength regions included in the calculation and
the way the distribution of SNR is treated across wavelengths are
major sources of uncertainty. In our coolest example ($T = 2500$\,K),
RV photon limits differ between our model calculations and empirical
values in the range 700--900\,nm. Model spectra do show an
overabundance of molecular absorption in comparison to observed
spectra \cite[see Appendix A in][]{2018A&A...612A..49R}, but since the
empirical RV estimates rely on stars that are very faint, they may
also partly be affected by read-out noise in this wavelength range.

From the information shown in Fig.\,\ref{fig:rvlambda}, we create a
grid of RV precision values for dwarf stars of spectral types F--M. We
adopt empirical values where available from CARMENES and/or HARPS, and
we rely on predictions from PHOENIX spectral models otherwise. We have
no empirical information about the $K$-band from CARMENES and assume
that in M dwarfs ($T < 4000$\,K) the spectral information content of
the $K$-band is similar to the $H$-band
\citep[see][]{2011A&A...532A..31R, 2016A&A...586A.101F,
  2018AJ....155..198A}. For hotter stars, we rely on our modelling
results.  At wavelengths and temperatures for which CARMENES and HARPS
information are both available, we take the average of the two. As
mentioned earlier, we generally scale RV photon noise based on stellar
$V$-magnitudes. For M dwarfs we also provide values based on
$J$-magnitudes. The grid values are summarized in
Tables\,\ref{tab:precision} and \ref{tab:precisionJ}.

\begin{deluxetable*}{rrrrrrrrrrr}
  \tablecaption{Grid of radial velocity precisions for dwarf stars at
    visual and infrared wavelengths. Values are given for an SNR of
    100 at $\lambda = 550$\,nm ($V$-band) and $R =
    100,000$.\label{tab:precision}}
  \tablecolumns{5} 
  \tablewidth{0pt} 
  \tablehead{ \colhead{} & \multicolumn{10}{c}{Wavelength range (nm)}}
  \startdata
   &   380 &   450 &   550 &   650 &   750 &   850 &   950 &  1150 &  1425 &  1980 \\
 &   450 &   550 &   650 &   750 &   850 &   950 &  1150 &  1425 &  1780 &  2380 \\
\hline
\colhead{$T_{\rm eff}$} & \multicolumn{10}{c}{RV photon noise (m s$^{-1}$)} \\
\hline
 7000 &  33.08 &  52.75 & 115.48 & 138.96 & 148.73 & 100.72 & 110.38 &  94.35 & 117.58 & 168.04 \\
 6000 &   1.86 &   1.45 &   2.77 &   4.13 &   4.65 &   4.94 &   6.04 &   7.16 &   6.03 &  15.17 \\
 5000 &   1.37 &   0.88 &   1.52 &   2.31 &   2.75 &   3.16 &   3.27 &   4.74 &   3.64 &   6.53 \\
 4500 &   1.61 &   0.86 &   1.41 &   1.88 &   2.42 &   2.86 &   2.23 &   4.07 &   2.65 &   4.22 \\
 3700 &   5.55 &   0.93 &   0.87 &   0.52 &   1.04 &   2.30 &   2.21 &   3.55 &   1.85 &   1.85 \\
 3200 &   1.54 &   0.37 &   0.48 &   0.26 &   0.30 &   0.72 &   0.83 &   1.34 &   0.91 &   0.91 \\
 2800 &   0.52 &   0.20 &   0.61 &   0.17 &   0.13 &   0.21 &   0.20 &   0.31 &   0.26 &   0.26 \\
 2500 &   0.30 &   0.17 &   0.28 &   0.12 &   0.12 &   0.13 &   0.07 &   0.07 &   0.08 &   0.08 \\

  \enddata
\end{deluxetable*}

\begin{deluxetable*}{rrrrrrrrr}
  \tablecaption{Same as Table\,\ref{tab:precision} but for an SNR of
    150 at $\lambda = 1200$\,nm ($J$-band).\label{tab:precisionJ}}
  \tablecolumns{5} 
  \tablewidth{0pt} 
  \tablehead{ \colhead{} & \multicolumn{8}{c}{Wavelength range (nm)}}
  \startdata
  &   550 &   650 &   750 &   850 &   950 &  1150 &  1425 &  1980 \\
  &   650 &   750 &   850 &   950 &  1150 &  1425 &  1780 &  2380 \\
  \hline
  \colhead{$T_{\rm eff}$} & \multicolumn{8}{c}{RV photon noise (m s$^{-1}$)} \\
  \hline
  3700 &    1.49 &   0.98 &   1.98 &   4.37 &   4.20 &   6.75 &   3.51 &   3.51 \\
  3200 &    1.95 &   0.92 &   1.09 &   2.57 &   2.98 &   4.77 &   3.27 &   3.27 \\
  2800 &    5.18 &   1.42 &   1.11 &   1.75 &   1.72 &   2.63 &   2.24 &   2.24 \\
  2500 &         &   3.69 &   3.50 &   3.98 &   2.12 &   1.96 &   2.32 &   2.32 \\
  \enddata
\end{deluxetable*}

\begin{figure*}
\centering
\includegraphics[width=.8\textwidth, bb = 20 245 590 450, clip=]{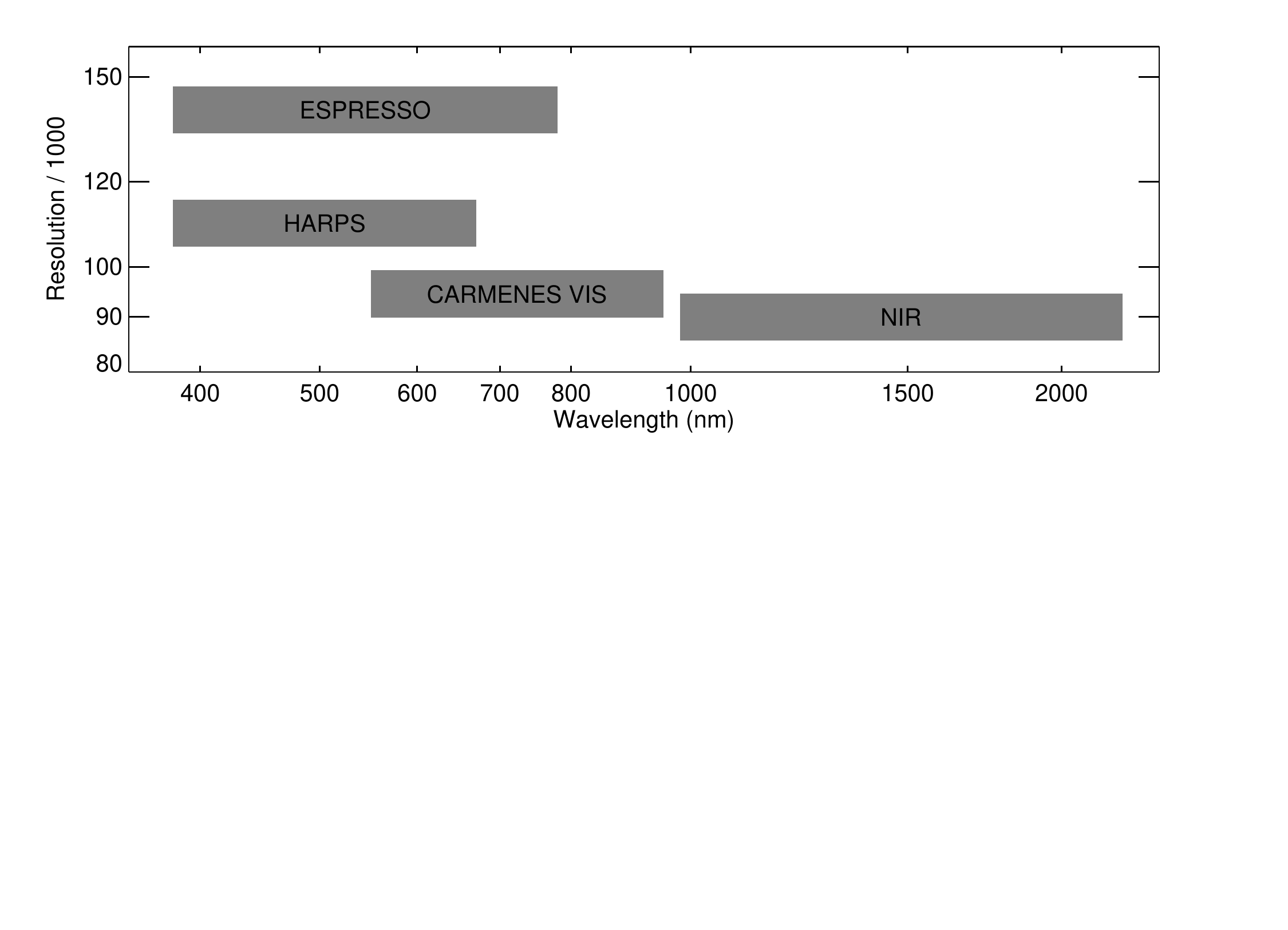}
\caption{\label{fig:spectrographs}Visualization of resolution and
  wavelength coverage of the four spectrograph designs used as
  examples.}
\end{figure*}

\section{Radial velocity photon noise catalogue}
\label{sect:catalogue}

For the stars in the catalogues introduced in
Sect.\,\ref{sect:sample}, we compute in this Section individual RV
photon noise limits expected in a typical observation for a few
typical spectrograph designs. In Sect.\,\ref{sect:hablim}, we estimate
minimum masses of planets detectable in the stars' liquid water
habitable zones following from the RV photon noise. In addition to our
catalogue, we provide an online tool with the relevant formulae. The
user can compute RV precision limits for individual stars or star
lists and for specific observational
configurations.\footnote{\url{http://www.astro.physik.uni-goettingen.de/research/rvprecision}}

\subsection{Spectrographs}

We compute RV precision estimates for four example spectrograph designs; (1)
an ultra-high resolution spectrograph at an 8-m telescope covering a very wide
wavelength range like ESPRESSO \citep{2014AN....335....8P}, (2) the
HARPS-design \citep{2003Msngr.114...20M} with very high resolution at visual
wavelengths operating at a 4m-class telescope, (3) a red optical design like
CARMENES-VIS \citep{2016SPIE.9908E..12Q} fed by a 4m telescope, and (4) a
near-infrared (NIR) design covering wavelengths redward of 900\,nm like, e.g.,
CARMENES-NIR \citep{2016SPIE.9908E..12Q}, SPIROU \citep{2017haex.bookE.107D},
NIRPS \citep{2017SPIE10400E..18W}, and GIANO \citep{2014SPIE.9147E..1EO}. The
parameters are summarized in Table\,\ref{tab:spectrographs} and visualized in
Fig.\,\ref{fig:spectrographs}. We specify instrument efficiency in terms of
SNR relative to HARPS and CARMENES-VIS values; these instruments deliver
comparable SNR after identical exposure times on objects of the same
brightness. Estimates for ESPRESSO and HARPS instrument throughput are based
on ESO exposure time calculators.\footnote{\url{http://eso.org/observing/etc}}
CARMENES-VIS estimates are taken from \cite{2018A&A...612A..49R}. Our NIR
design SNR estimate is based on a SPIROU estimate \citep{2017haex.bookE.107D},
which is somewhat higher than the CARMENES-VIS and HARPS throughput. For our
calculations, we assume an exposure time of 5\,min for ESPRESSO and 10\,min
for the other spectrograph designs. This is an arbitrary choice motivated by
typical exposures used at 4m- and 8m-class telescopes for high-precision RV
observations. In exposures shorter than a few minutes, solar-like oscillations
likely do not average out, in particular in higher-mass stars
\citep{2019AJ....157..163C}. For longer exposure times, RV precision scales
with $\sqrt{t_{\rm exp}}$, but exposures should not be longer than a few ten
minutes to avoid large systematic errors from barycentric motion
\citep[e.g.,][]{2019MNRAS.489.2395T}.

\begin{deluxetable*}{lcccc}
  \tablecaption{Spectrograph parameters as used for the RV precision
    estimates. \label{tab:spectrographs}}
  \tablecolumns{5} 
  \tablewidth{0pt} 
  \tablehead{\colhead{Instrument} & \colhead{$\lambda$ (nm)} & \colhead{$R$} & \colhead{rel.\,SNR} & \colhead{Exp. time}}
  \startdata
  ESPRESSO     & 380 --  780 & 140,000 &  2  & 5 min\\
  HARPS        & 380 --  670 & 110,000 &  1  & 10 min\\
  CARMENES VIS & 550 --  950 &  94,600 &  1  & 10 min\\
  NIR          & 980 -- 2440 &  90,000 & 1.2 & 10 min\\
  \enddata
\end{deluxetable*}

\subsection{RV photon noise}
\label{sect:rvprecision}

\begin{figure*}
\gridline{\fig{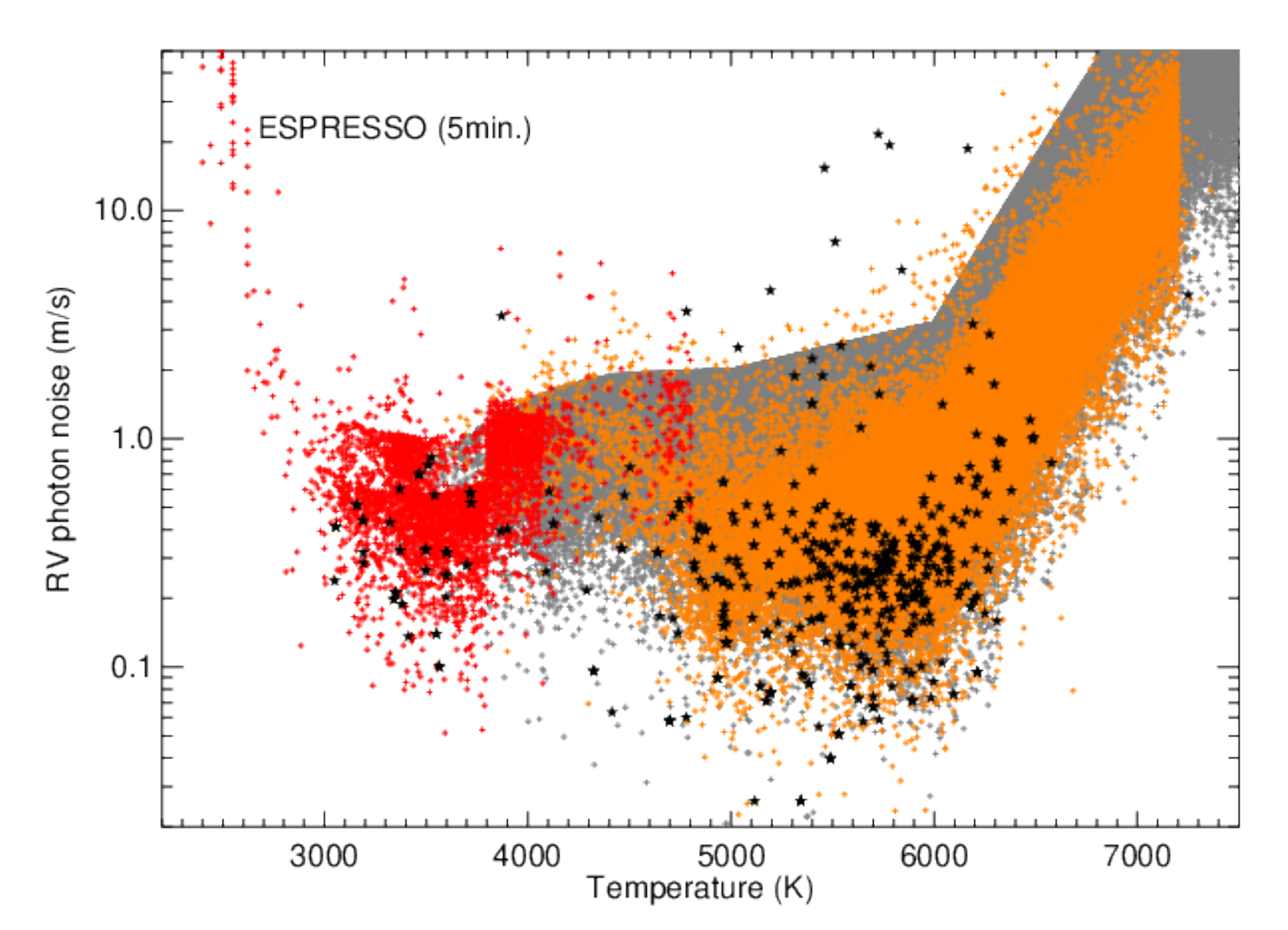}{0.5\textwidth}{(a)}
          \fig{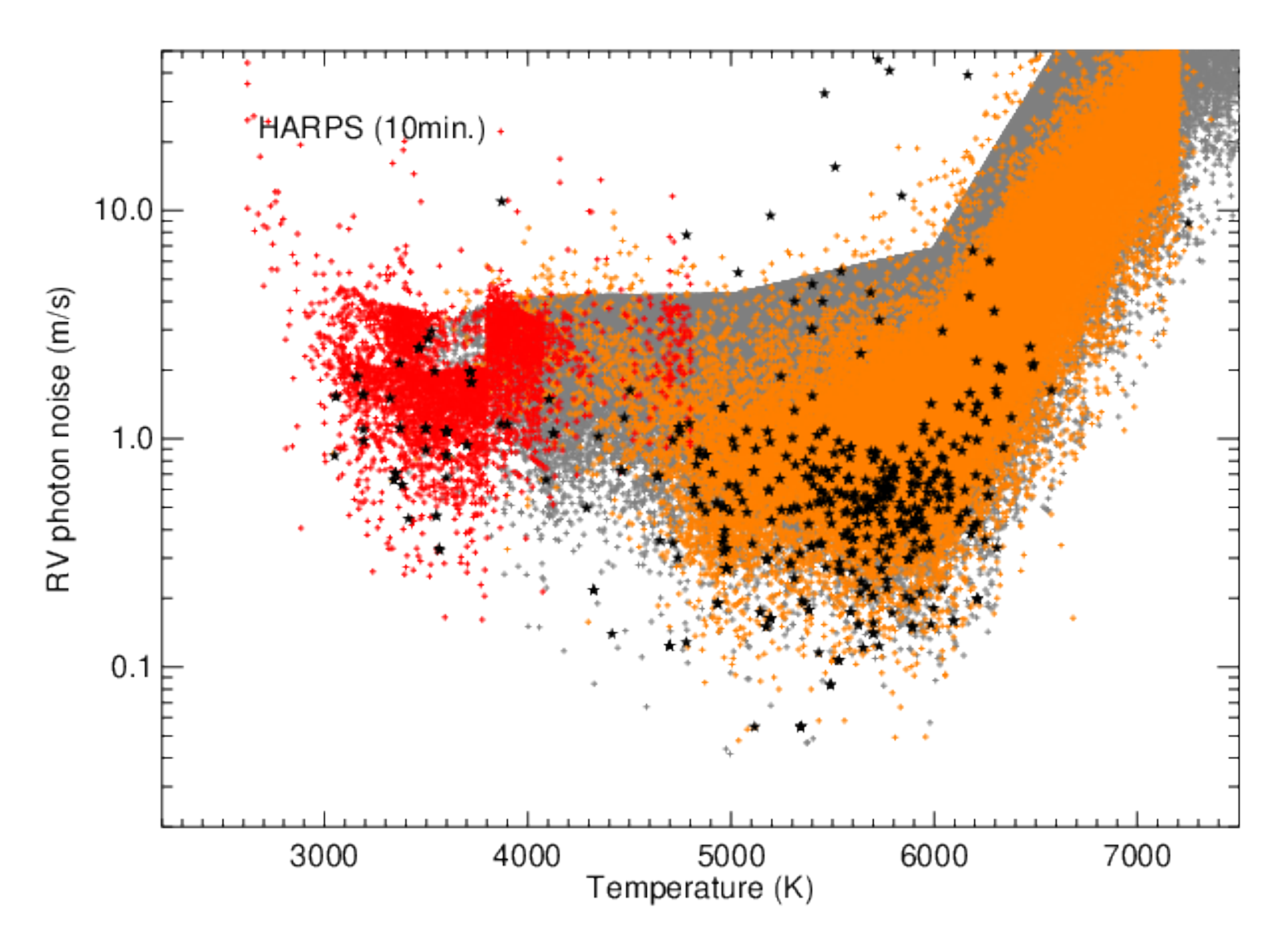}{0.5\textwidth}{(b)} }
\gridline{\fig{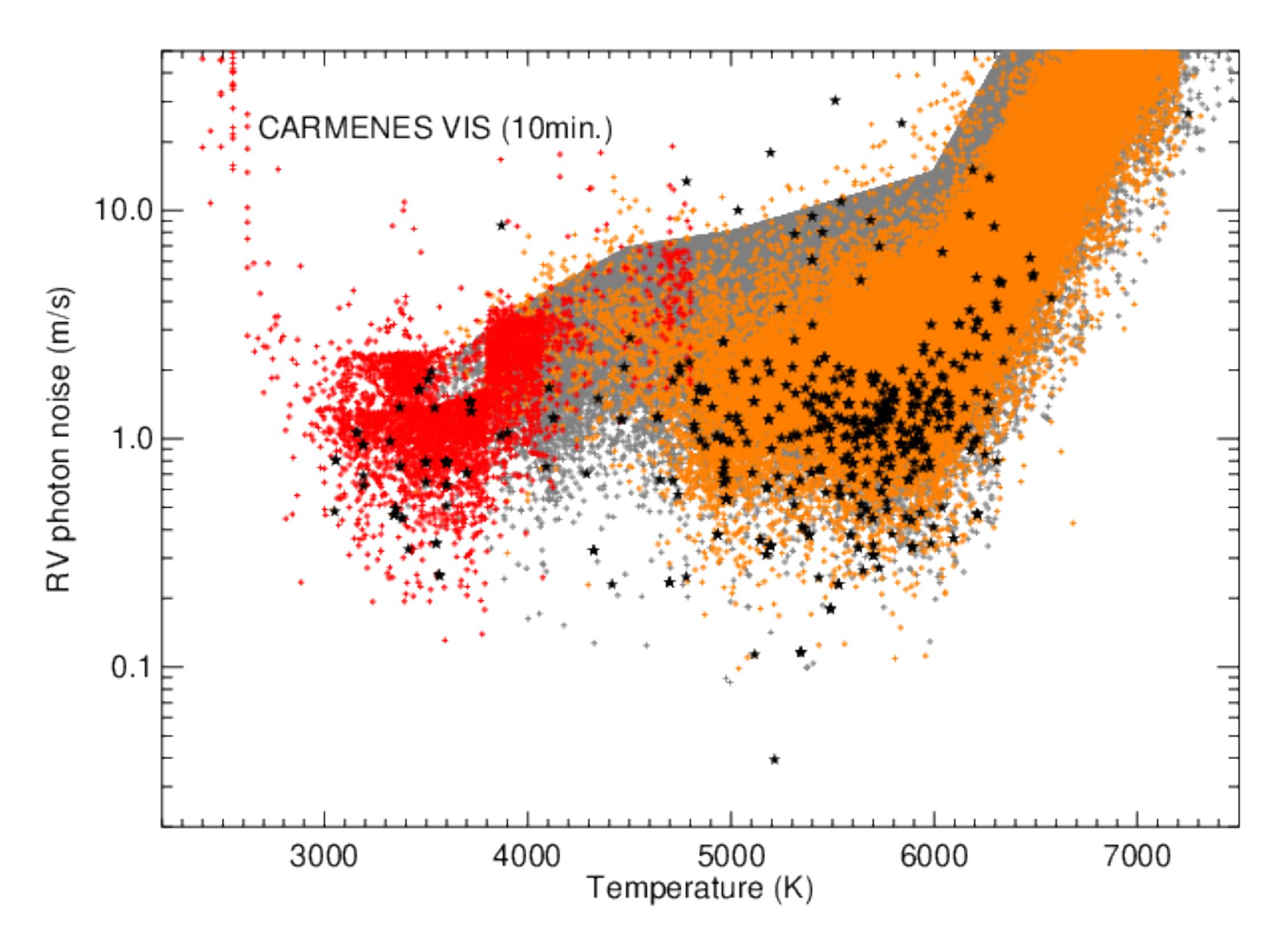}{0.5\textwidth}{(c)}
          \fig{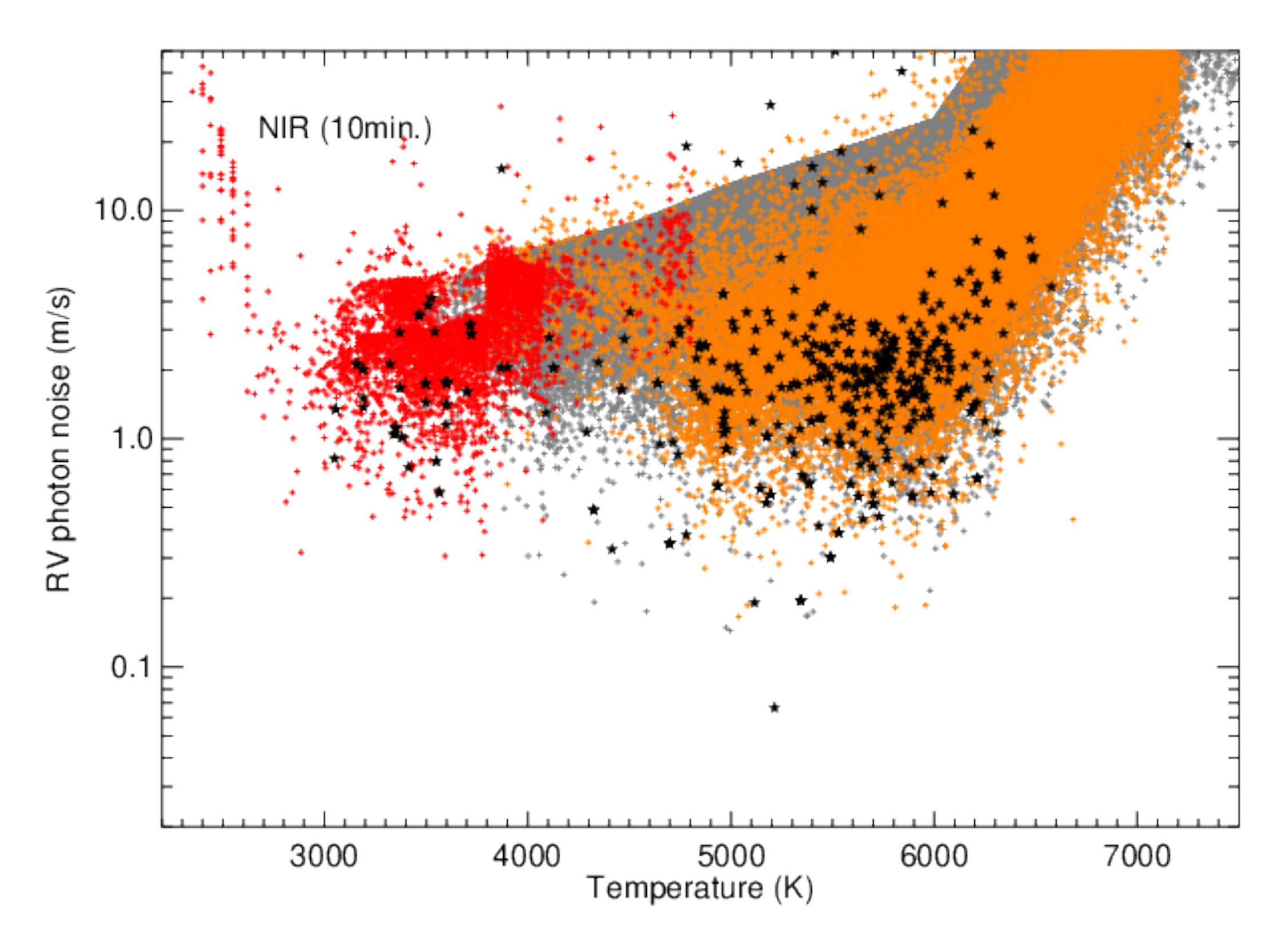}{0.5\textwidth}{(d)}
          }
          \caption{\label{fig:rvprecisions}RV photon noise for the
            stars of our sample estimated for 10\,min observations
            with the spectrograph designs (a) ESPRESSO, (b) HARPS,
            (c) CARMENES VIS, and (d) NIR, see
            Table\,\ref{tab:spectrographs}. Colors indicate the
            reference catalogues as in Fig.\,\ref{fig:HRD}.}
\end{figure*}

In order to estimate the RV photon noise for the stars of our sample,
we scale the SNR according to stellar brightness (and RV photon noise
$\propto$ 1/SNR). For stars from the F--K dwarf catalogues, we scale
SNR according to $V$-band magnitudes (Table\,\ref{tab:precision}). For
stars from Gaia DR2 we assume Gaia-$g = V$. For stars cooler than
3800\,K, we do not rely on the $V$-band magnitudes alone because these
stars are very faint at optical wavelengths and not all catalogues
provide $V$-magnitude; we use the $J$-magnitudes instead
(Table\,\ref{tab:precisionJ}). If the $V$-magnitude is available as
well, we use it to calculate RV precision at wavelengths shorter than
550\,nm. If no $V$-magnitude is available, we neglect this wavelength
range. In our SNR calculations, we include read-out noise of 20
electrons per extracted pixel in the reduced spectrum. This value can
differ a lot for different spectrographs, for example spectrographs
with image or pupil slicers typically collect more read-out noise than
others. Read-out noise can be important at optical wavelengths in
low-mass stars, where spectral information is very dense and
relatively low values of SNR can still provide high RV precision. For
simplicity, we use the term \emph{RV photon noise} although our
results can be read-out noise limited in very faint stars.

The RV photon noise achievable with the four fiducial spectrograph
models are shown in Fig.\,\ref{fig:rvprecisions}. We re-emphasize that
these values are theoretical lower limits that are often well below
the instrumental or stellar jitter. We also recall that these values
are estimates valid for fixed exposure times of 5 and 10\,min
(Table\,\ref{tab:spectrographs}), and that lower photon noise limits
can always be achieved through longer exposures. While the absolute
values for our example observations can be scaled by adjusting
exposure times, the relative performances between stars of different
temperatures, and the comparison between instruments of different
design, are fixed within the limits of throughput, performance
limitations, and uncertainties in our RV noise estimates above as well
as its scaling with stellar brightness including a read-out
noise-floor. It is therefore very instructive to investigate the
temperature-dependent performance of different spectrograph
designs. Finally, we note that all our calculations are carried out
for the full set of stars regardless of their position on the sky,
i.e., they are valid for instruments with similar performance as our
spectrographs examples, but not all of the objects are observable from
the locations of actually existing instruments.

The distribution of RV photon noise in Fig.\,\ref{fig:rvprecisions}
shows the consequences from the choice of wavelength coverage. The
ESPRESSO design covers the entire optical spectrum between 380\,nm and
780\,nm.  Among the stars available in our catalogues (ca.\
3000--7000\,K), many of the brightest stars can be observed with RV
photon noise limits around 10\,cm\,s$^{-1}$ or better.  After 5\,min,
the ESPRESSO design can collect enough photons to reach an RV noise
limit better than 2\,m\,s$^{-1}$ for almost all stars in our
catalogues except for the coolest (and faintest) ones and the hottest
stars ($> 6000$\,K) that are likely rotating too rapidly. Note again
that we did not include actual measurements of $v\,\sin{i}$ here (see
above). The RV photon limits for the other three spectrographs are
generally a bit higher because our fiducial ESPRESSO exposures collect
about a factor of two more photons (at the same wavelengths).

Like ESPRESSO, the HARPS design also has a spectral resolution well
above $R=10^5$ and covers wavelengths redward of 380\,nm, but the
cutoff at long wavelengths occurs approximately 110\,nm bluer than for
the ESPRESSO format. The missing red wavelength range carries a lot of
RV information in M dwarfs \citep{2018A&A...612A..49R}, and its lack
is the reason why the HARPS design performs worse in M dwarfs relative
to hotter stars in comparison to the ESPRESSO design. The sweet spot
of the HARPS design is in the G-dwarfs where some targets provide
enough information for RV measurements at the 10\,cm\,s$^{-1}$ level
after 10\,min. Furthermore, also a large number of M dwarfs can be
observed with RV photon noise limits better than 1\,m\,s$^{-1}$ after
the same exposure times, which is consistent with the results of,
e.g., \citet{2009A&A...507..487M, 2013A&A...549A.109B,
  2013A&A...556A.126A, 2014MNRAS.443L..89A, 2016Natur.536..437A}.

The two other spectrograph designs in our comparison, the CARMENES VIS
and the NIR design, are both optimized for M dwarfs. Their RV photon
limits are skewed with respect to the curves from the optical designs
like ESPRESSO and HARPS. Dwarfs later than spectral type M5/M6
($\sim$2800\,K) become drastically fainter towards lower temperature,
with the result that the SNR of individual exposures becomes very
low. Thus, for the conditions simulated here, the RV photon limits in
late-M dwarfs ($<$2800\,K) for observations at our 4m-class designs
are not better than a few m\,s$^{-1}$ even in the closest
objects. Nevertheless, at this temperature, the wavelength coverage of
the CARMENES VIS and NIR designs provide significant improvement with
respect to the visible-light designs. For the brightest late-M dwarfs
on the sky, RV photon limits better than 10\,m\,s$^{-1}$ can be
reached within 10\,min. Again, we want to emphasize that our RV photon
limits are theoretical values that do not take into account stellar
jitter and individual values of rotational broadening. Mid- and late-M
dwarfs are often rapid rotators with high values of projected rotation
velocities, $v\,\sin{i}$, as large as a few ten km\,s$^{-1}$
\citep[e.g.,][]{2003ApJ...583..451M, 2012AJ....143...93R}. For the
selection of targets in an RV survey, the effect of rotational
broadening should be taken into account.

\subsection{Optical or near-IR?}

\begin{figure}
\centering
\includegraphics[width=\textwidth]{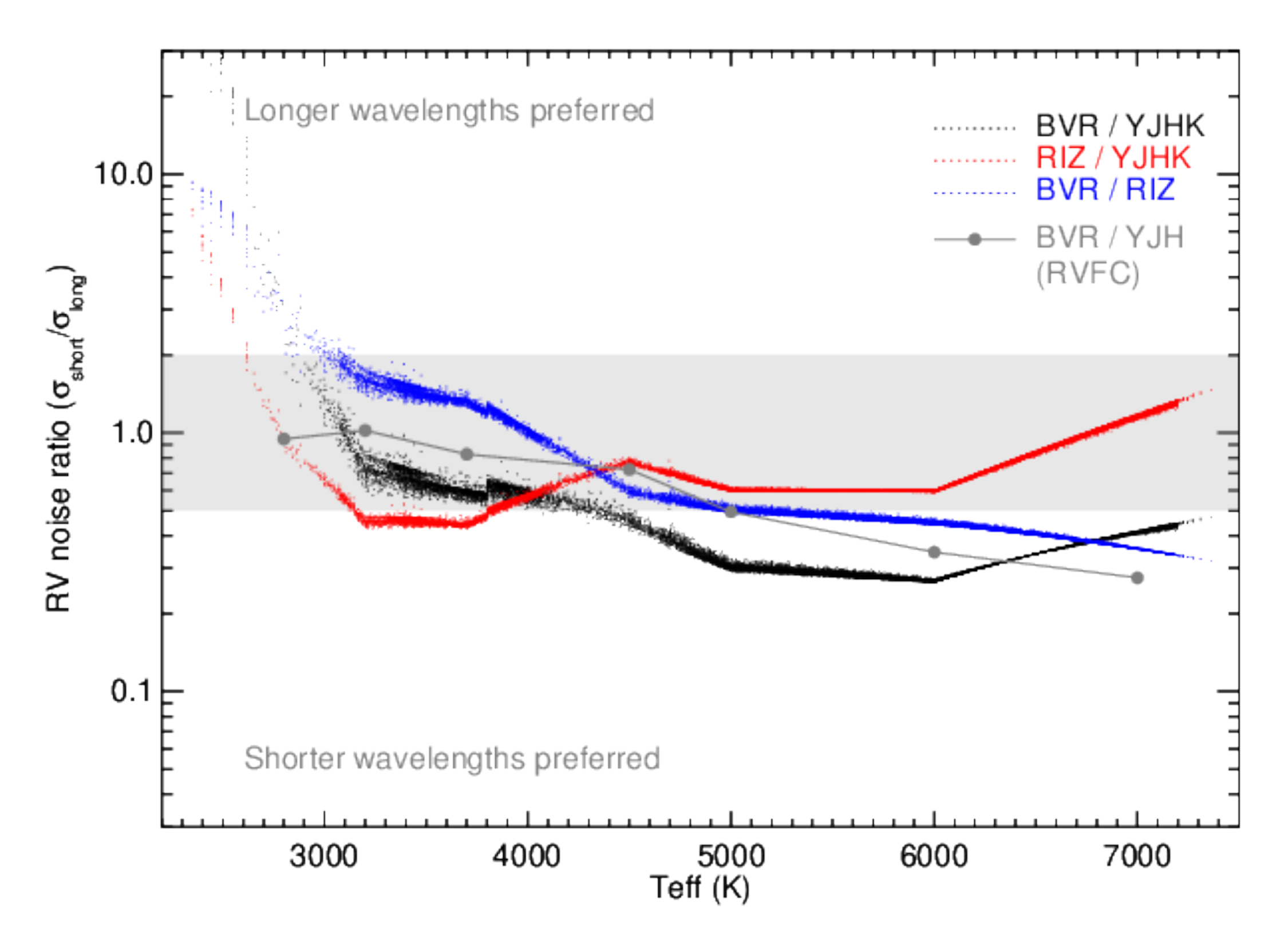}
\caption{\label{fig:ratios}Ratios between RV photon noise achieved in
  observations of dwarf stars with different spectrograph designs as a
  function of stellar effective temperature. For each star, black
  points shows the ratio between the performance of a HARPS-like
  design covering the BVR-bands and a near-IR SPIROU-like design
  covering YJHK. The red and the blue points show RV noise ratios
  between a red-optical instrument covering RIZ (like CARMENES-VIS) and
  the near-IR design, and between the optical and red-optical designs,
  respectively. The grey circles show results from calculations using
  the radial velocity follow-up calculator from
  \citet{2018AJ....156...82C}. Within uncertainties, two spectrographs
  perform similarly if the ratio is located in the range 0.5--2.0
  (grey).}
\end{figure}

With our RV noise calculations, we can compare performances of
spectrographs designed for different wavelengths.
\citet{2018AJ....156...82C} quantified the observational effort
required for RV measurements in TESS targets. They compared an optical
HARPS-like design to a near-infrared design covering wavelengths
accross spectral bands $YJH$. An important conclusion from their work
is that M dwarfs with $T \la 3800$\,K are more efficiently observed
with near-IR spectrographs than with optical ones. Here, we want to
revisit the question whether optical or near-IR spectrographs are more
efficient for M-dwarf RV follow-up.

In Fig.\,\ref{fig:ratios}, we show for each star the ratios between
photon noise limits achieved with different spectrograph designs as a
function of stellar temperature. In our comparison, we include the
optical HARPS-like design covering the $BVR$ bands, the CARMENES-VIS
design covering the $RIZ$ bands, and the near-IR design covering
$YJHK$. Ratios between the RV noise limits are independent of other
design details assuming that all parameters except wavelength coverage
are comparable.

The comparison between optical ($BVR$) and near-IR ($YJHK$)
instruments are shown in black in Fig.\,\ref{fig:ratios}. In sun-like
stars, the optical setup outperforms the near-IR one by roughly a
factor of four. High values of SNR and the much higher density of
spectral features in the blue part of the spectrum favor the
instrument design covering shorter wavelengths. In stars of cooler
temperature, according to our calculations, the near-IR design is more
efficient than the optical design at stellar temperatures $T \la
3200$\,K. This result is significantly different from the breakpoint
at 3800\,K reported by \citet{2018AJ....156...82C} with potentially
relevant consequences for RV follow-up projects. It is therefore
important to understand the reasons behind this
discrepancy. \citet{2018AJ....156...82C} compared the total
observational effort to characterize potential TESS planets. In
addition to photon noise, they included other sources of RV jitter,
but they chose to not introduce any wavelength dependence in their
jitter terms. This means that their comparison of RV noise limits
between spectrographs is conceptually not different to ours. The
calculations of \citet{2018AJ....156...82C} were based on the same
synthetic model spectra as our model calculations, but they applied
the empirical correction factors \citet{2018AJ....155..198A}
determined for Barnard's star (M4) to all their calculations. For a
more direct comparison between our ratios and the results from
\citet{2018AJ....156...82C}, we use their online
calculator\footnote{\url{http://maestria.astro.umontreal.ca/rvfc/}} to
calculate RV photon noise limits for seven stars with typical
main-sequence parameters. RV noise limits are calculated for the HARPS
(optical) and SPIROU (near-IR) spectrograph designs as offered in the
tool. We plot the ratios $\sigma_{\rm RV, opt} / \sigma_{\rm RV,
  nearIR}$ as grey circles in Fig.\,\ref{fig:ratios}. We find that
these ratios are very different from the estimated observation efforts
reported in Fig.\,3 of \citet{2018AJ....156...82C}. Specifically, our
calculations predict that the optical and near-IR designs are
performing relatively similar in dwarfs stars with $T \la 5000$\,K. In
hotter stars, the optical design outperforms the near-IR by about a
factor of three. The near-IR design is not predicted to be
significantly more efficient than the optical at stellar temperatures
$T \ge 2800$\,K, which is the lower temperature limit of the online
tool. We suspect that the discrepancy between the RV efficiency
estimated in \citet{2018AJ....155..198A} and our results from their
online tool is caused by read-out noise dominating the results in
\citet{2018AJ....155..198A}, predominantly in the $BVR$-bands. Their
sample includes many relatively faint objects while our seven test
objects where chosen to be bright enough to neglect read-out
noise. The effect of read-out noise can be seen in the RV noise ratios
of our (relatively bright) sample stars in Fig.\,\ref{fig:ratios}; the
(relatively small) scatter in the RV noise at a given temperature is
caused by read-out noise.

The optical and near-IR design both do not include the $I$-band, which is the
wavelength range carrying most RV information among M dwarfs
\citep[see][]{2018AJ....155..198A, 2018A&A...612A..49R}. We also compare the
red-optical design ($RIZ$) to the optical ($BVR$) and to the near-IR designs
($YJHK$) in Fig.\,\ref{fig:ratios}. It is not surprising that the $RIZ$-design
(red-optical) is more efficient than the $BVR$-design (optical) in stars with
$T_{\rm eff} \la 4000$\,K. For the comparison between the $RIZ$-design to the
$YJHK$-design (near-IR), one could expect that the redder $YJHK$-design
outperforms the bluer $RIZ$-design because of the higher flux density in the
near-IR. Nevertheless, as shown for example in \citet{2018A&A...612A..49R},
the much higher information content in the $I$-band, predominantly caused by
molecular lines, compensates for this effect. As a result, the red-optical
($RIZ$) design is more efficient than the near-IR ($YJHK$) in M dwarfs with
$T_{\rm eff} \ga 2700$\,K, i.e., M dwarfs earlier than approximately M7.

\subsection{Reaching the 1\,m\,s$^{-1}$ and 10\,cm\,s$^{-1}$ noise limits}

\begin{figure*}
\gridline{\fig{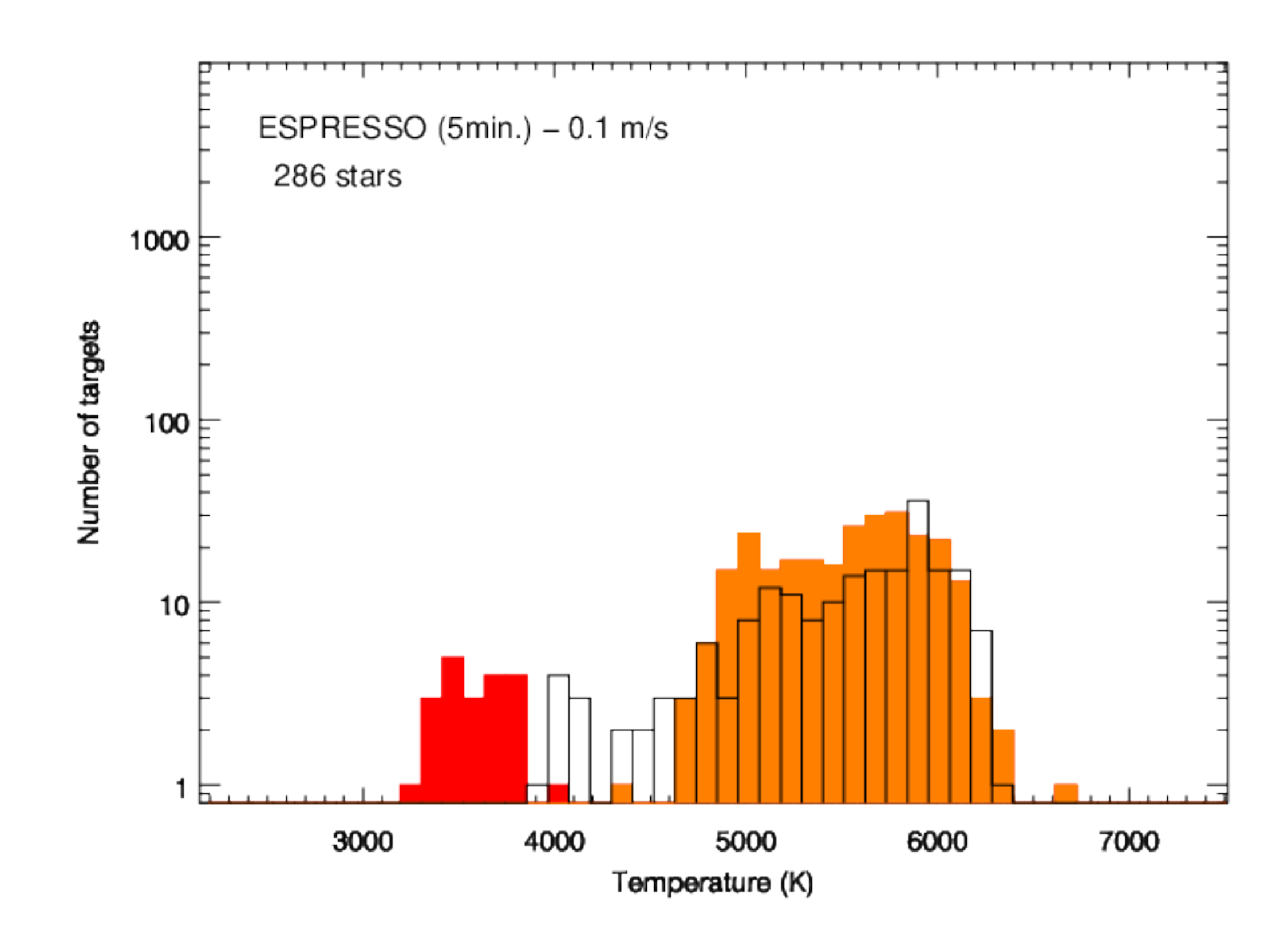}{0.5\textwidth}{(a)}
          \fig{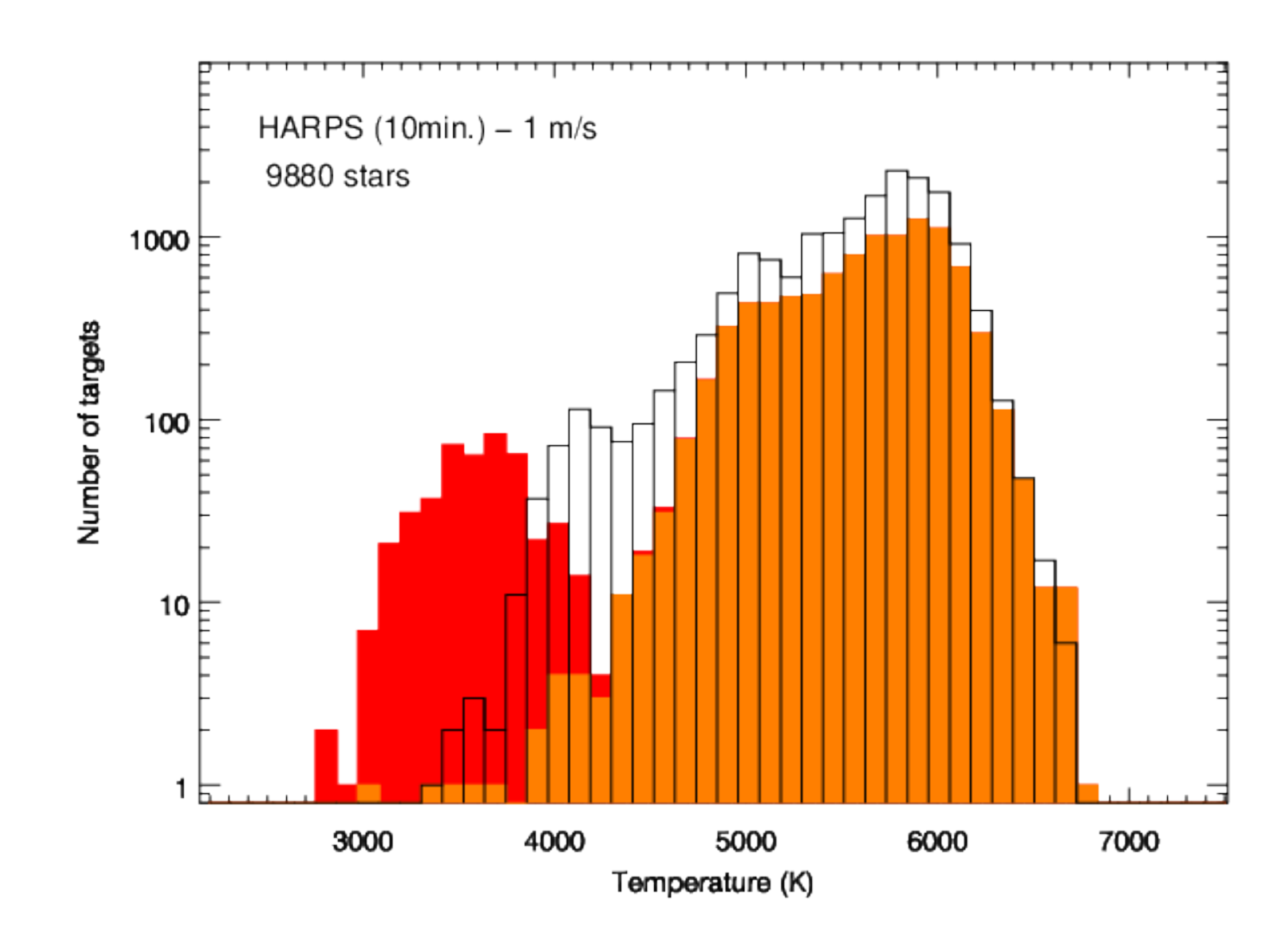}{0.5\textwidth}{(b)} }
\gridline{\fig{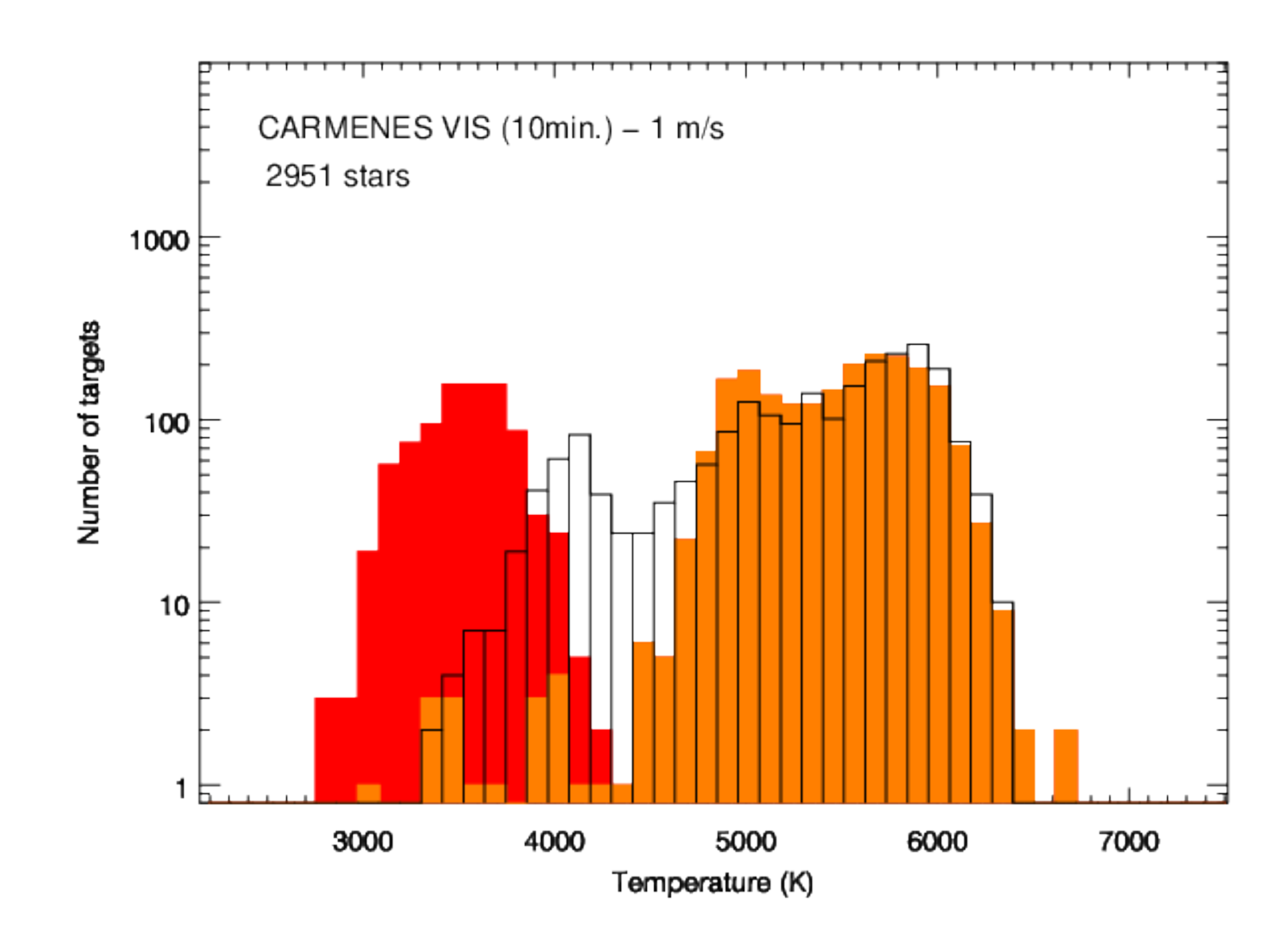}{0.5\textwidth}{(c)}
          \fig{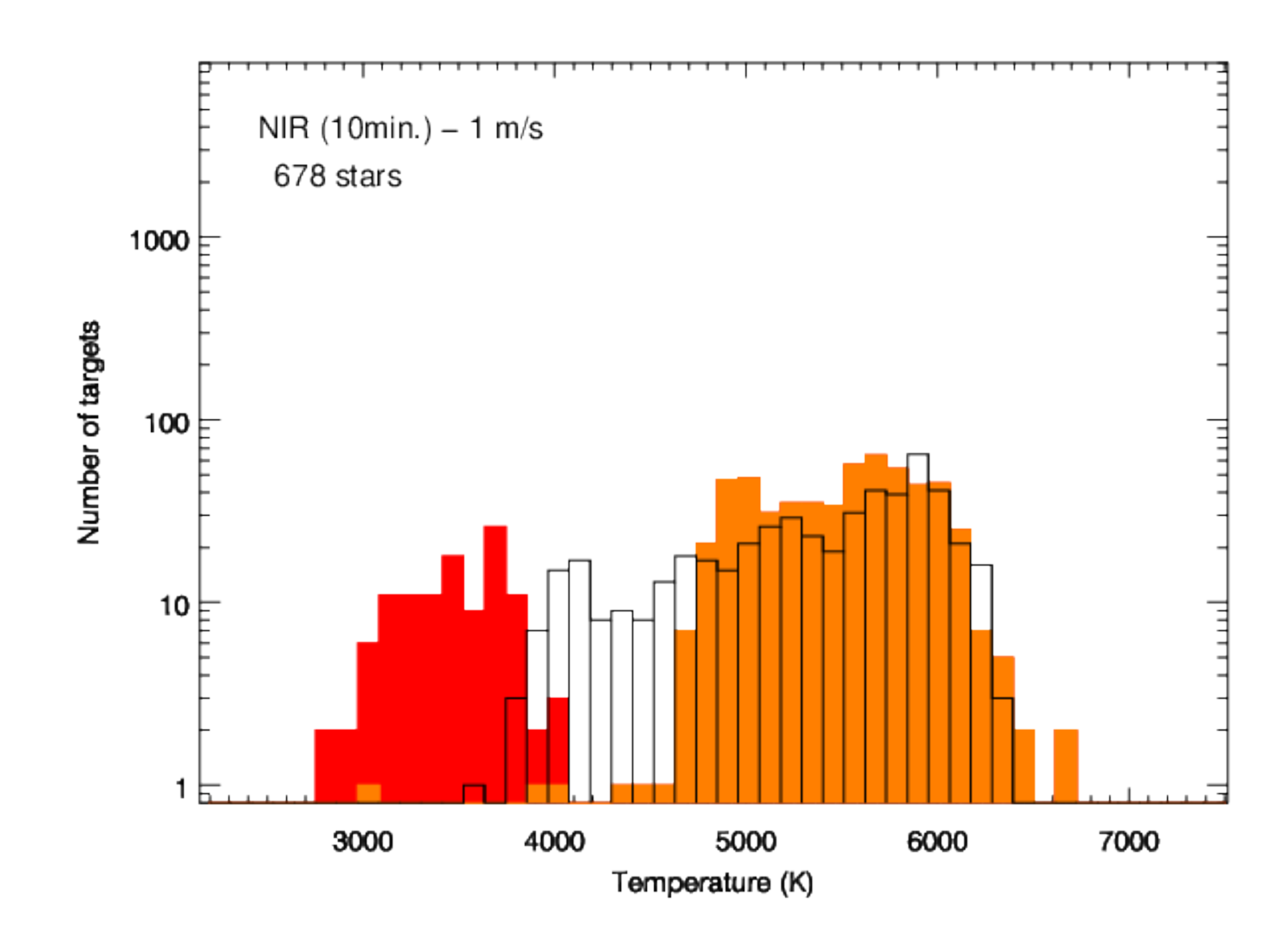}{0.5\textwidth}{(d)}
          }
          \caption{\label{fig:rv1mshisto}Histograms of the number of
            stars from the catalogues in Table\,\ref{tab:catalogues}
            for which an RV photon limit of 0.1\,m\,s$^{-1}$ (ESPRESSO
            design -- shown in panel a) or 1\,m\,s$^{-1}$ (panels
            b--d) can be reached within 10\,min. Colors indicate stars
            from different catalogues as in Fig.\,\ref{fig:HRD}.}
\end{figure*}

An RV precision of 1\,m\,s$^{-1}$ is often the goal of modern
spectrograph design, and ESPRESSO was designed with the goal of
measuring RV variations on the order of 0.1\,m\,s$^{-1}$ similar to
the effect of Earth on the Sun. As discussed above, the time required
to reach a given RV photon limit depends on the star's brightness, the
size and efficiency of the telescope, and the spectrograph design. For
the example designs of spectrographs we now ask the question for how
many stars from our sample a given RV photon limit can be reached
after a fixed exposure time.

We show in Fig.\,\ref{fig:rv1mshisto} histograms for the stars in
which an RV photon limit of 1\,m\,s$^{-1}$ (0.1\,m\,s$^{-1}$) can be
reached within 10\,min (5\,min) for the HARPS, CARMENES VIS, and NIR
designs (ESPRESSO design). For the ESPRESSO design, we find that 286
stars from our sample are candidates for observations at the
0.1\,m\,s$^{-1}$ level. Most of them are stars with temperatures
between 5000\,K and 7000\,K but there are also some ten stars cooler
than 4000\,K (spectral types early- to mid-M). Considering the
4m-class instruments, we find that with the HARPS design, 9880 stars
from our sample are generally suitable for observations to determine
RVs with a photon noise limit better than 1\,m\,s$^{-1}$ within
10\,min. Most of these stars are F- and G-type stars. The numbers
decrease towards cooler temperatures, and there are several 100 early-
to mid-M dwarfs that fulfill the criteria. The CARMENES VIS design is
less efficient for stars of spectral type F--K but more efficient in
very cool stars. This results in higher numbers of mid- and late-M
stars reaching the targeted RV precision. Overall, we find 2951 stars
in which a limit of 1\,m\,s$^{-1}$ can be reached in a 10\,min
exposure. For the NIR design, we find that the 1\,m\,s$^{-1}$ level
can be reached in 678 stars. For all cases, we find a lack of targets
between 4000\,K and 5000\,K (K-dwarfs). This type of stars is not well
covered by the catalogues we used. Gaia DR2 contains a number of stars
that are probably suitable for these measurements.

\section{Habitable zone limits}
\label{sect:hablim}

\begin{figure*}
\gridline{\fig{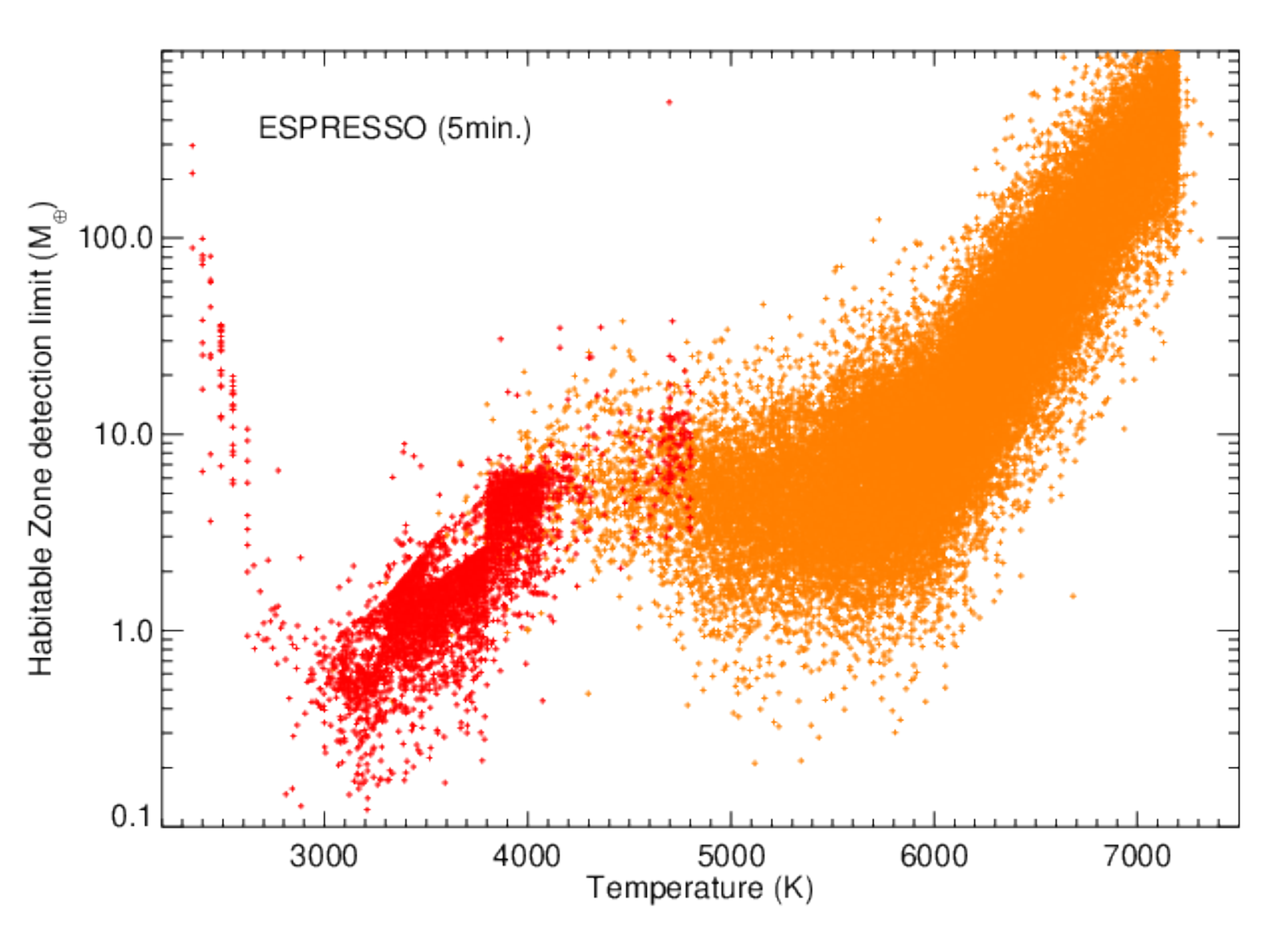}{0.5\textwidth}{(a)}
          \fig{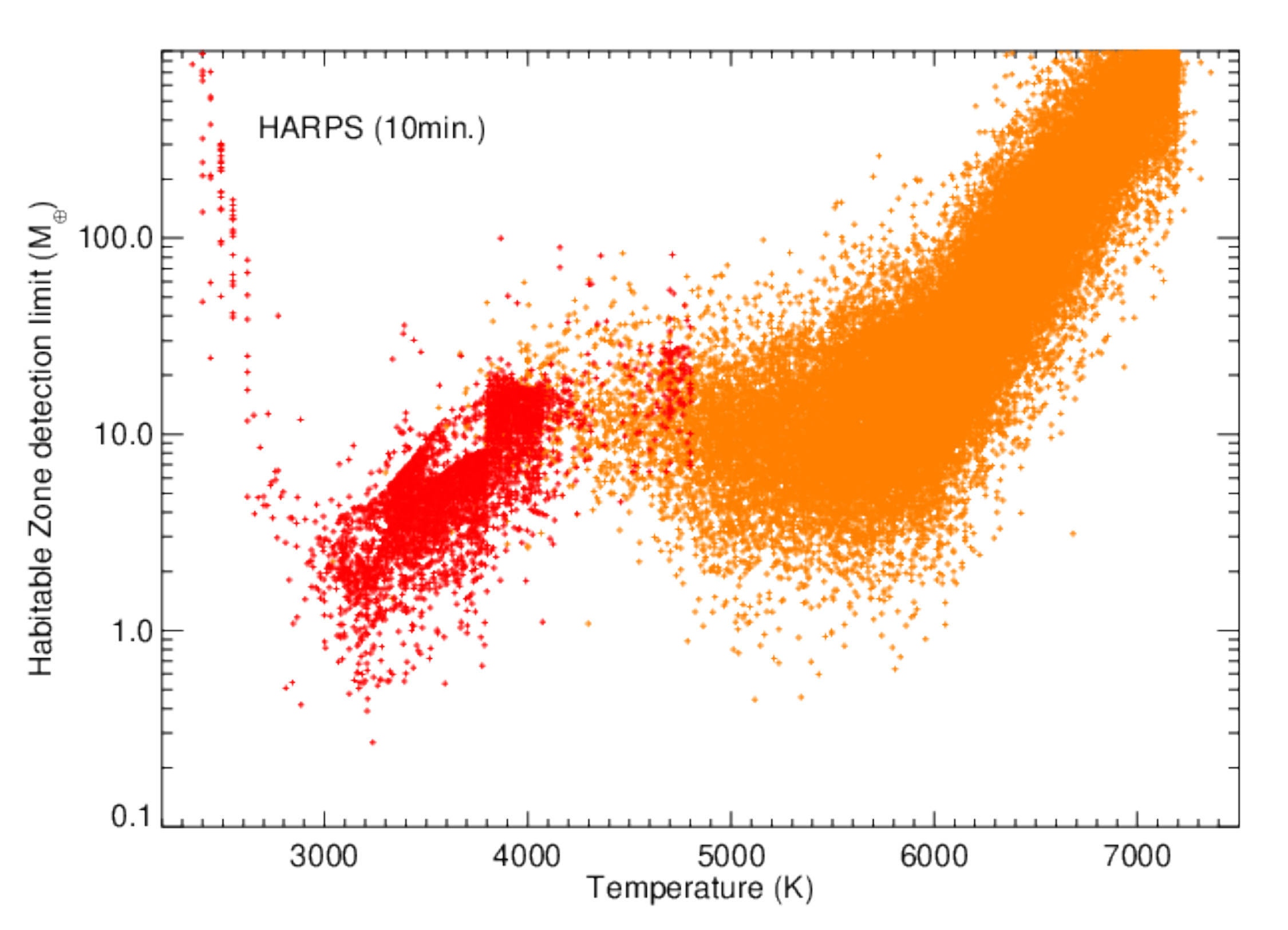}{0.5\textwidth}{(b)} }
\gridline{\fig{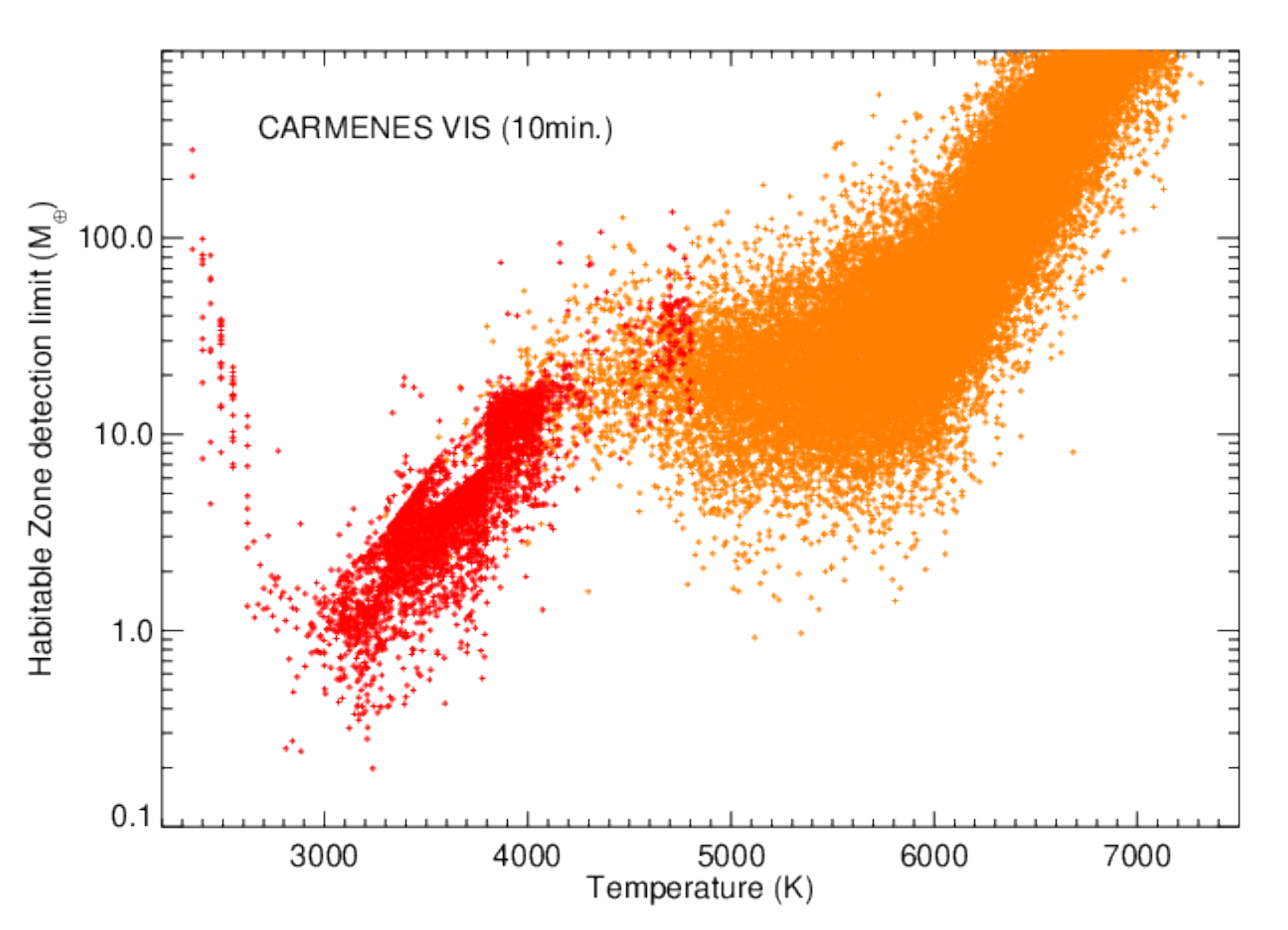}{0.5\textwidth}{(c)}
          \fig{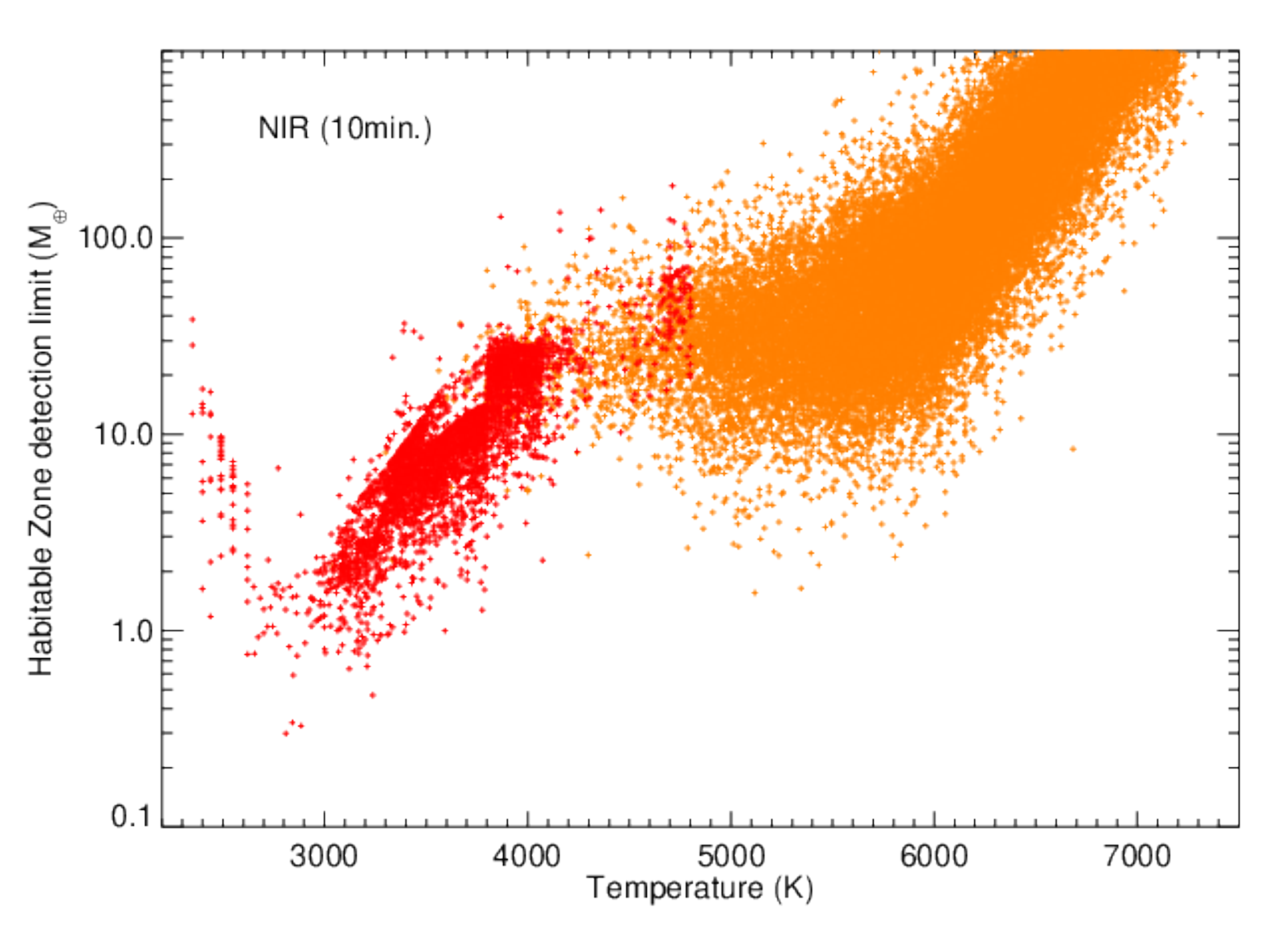}{0.5\textwidth}{(d)}
          }
          \caption{\label{fig:hzlimit}Minimum masses of planets
            orbiting inside the HZ of stars from our sample that can
            be detected with the four observational setups as
            explained in the text. Colors indicate stars from
            different catalogues as in Fig.\,\ref{fig:HRD}.}
\end{figure*}

\begin{figure*}
\gridline{\fig{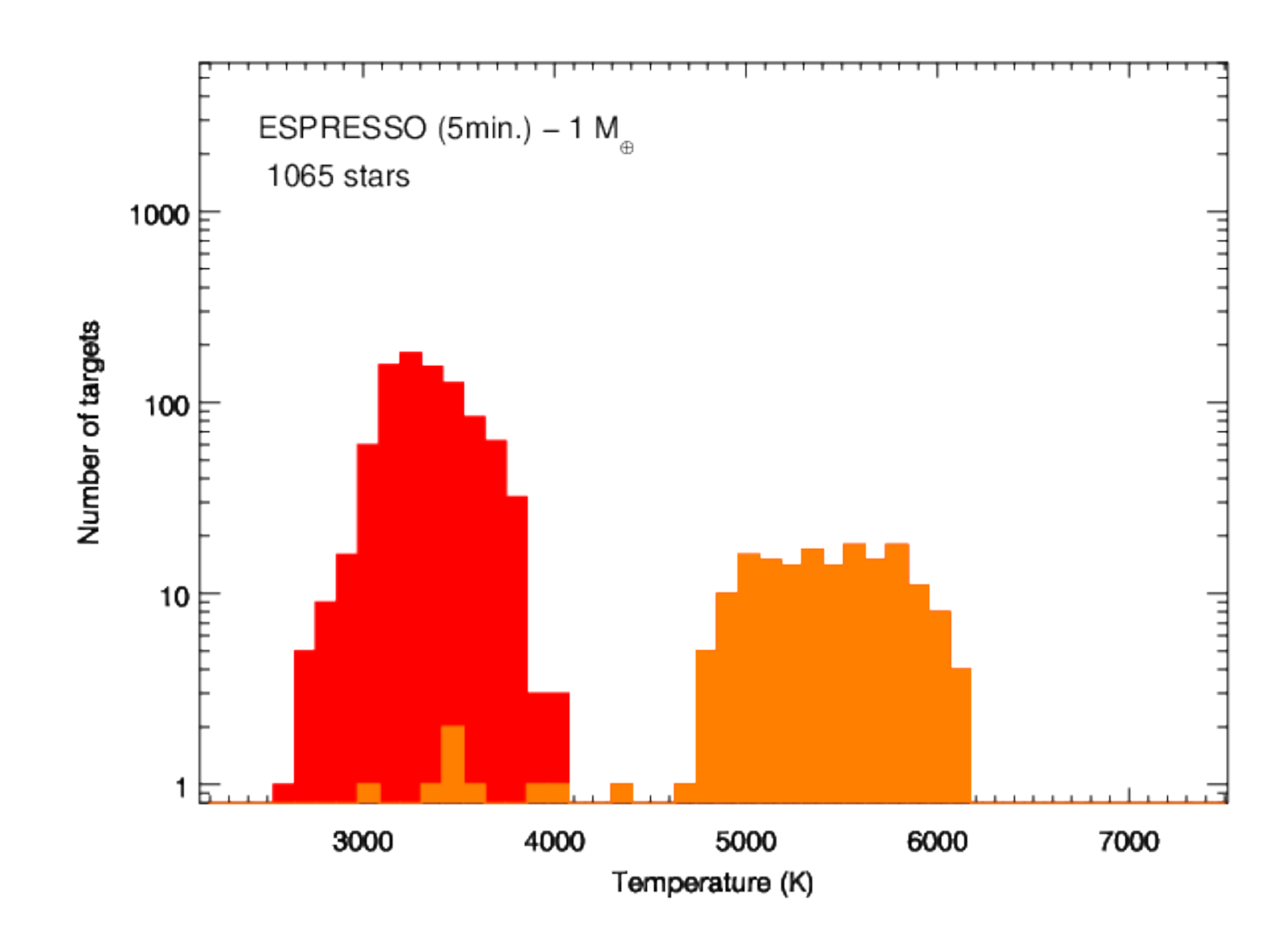}{0.5\textwidth}{(a)}
          \fig{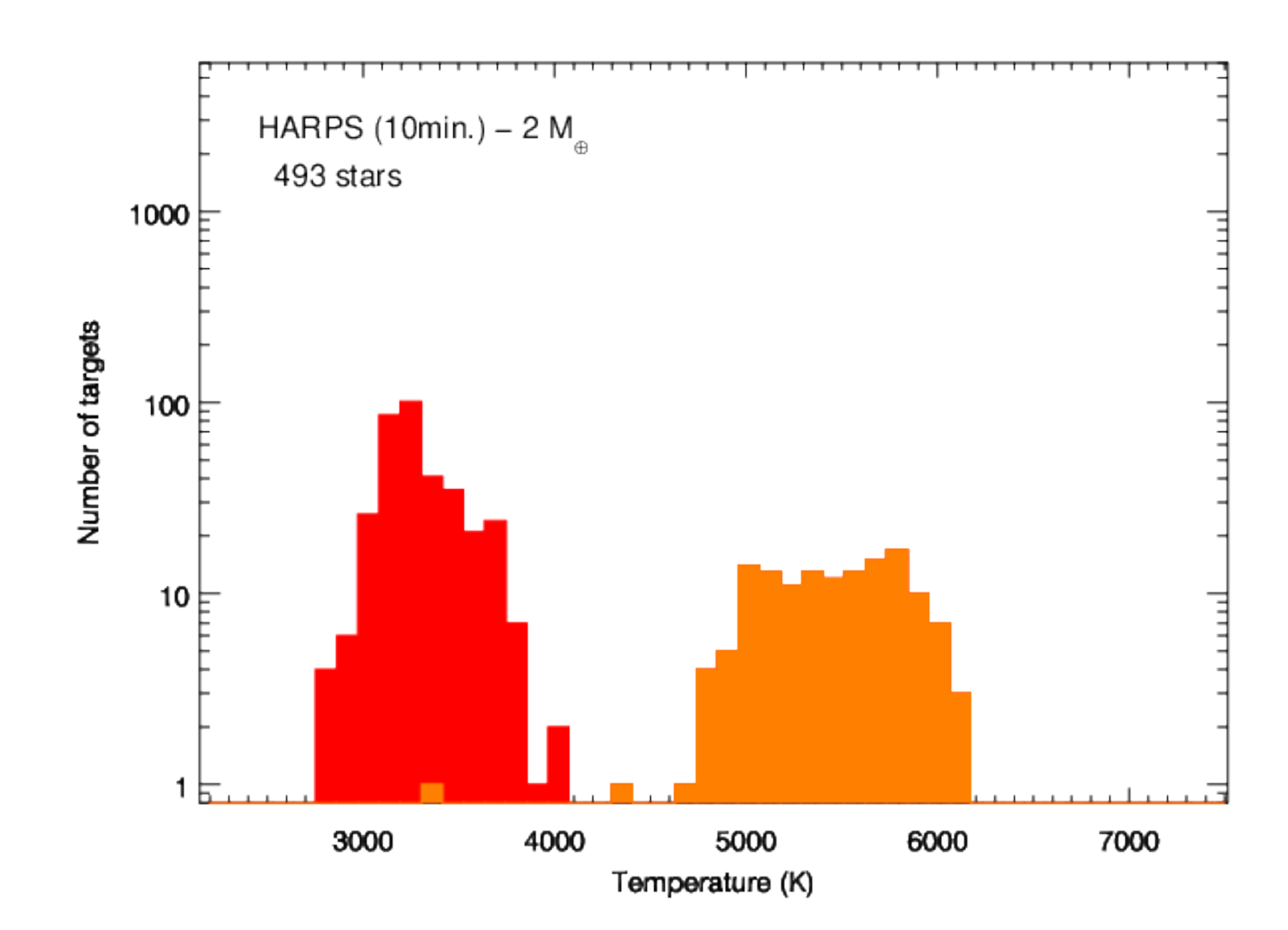}{0.5\textwidth}{(b)} }
\gridline{\fig{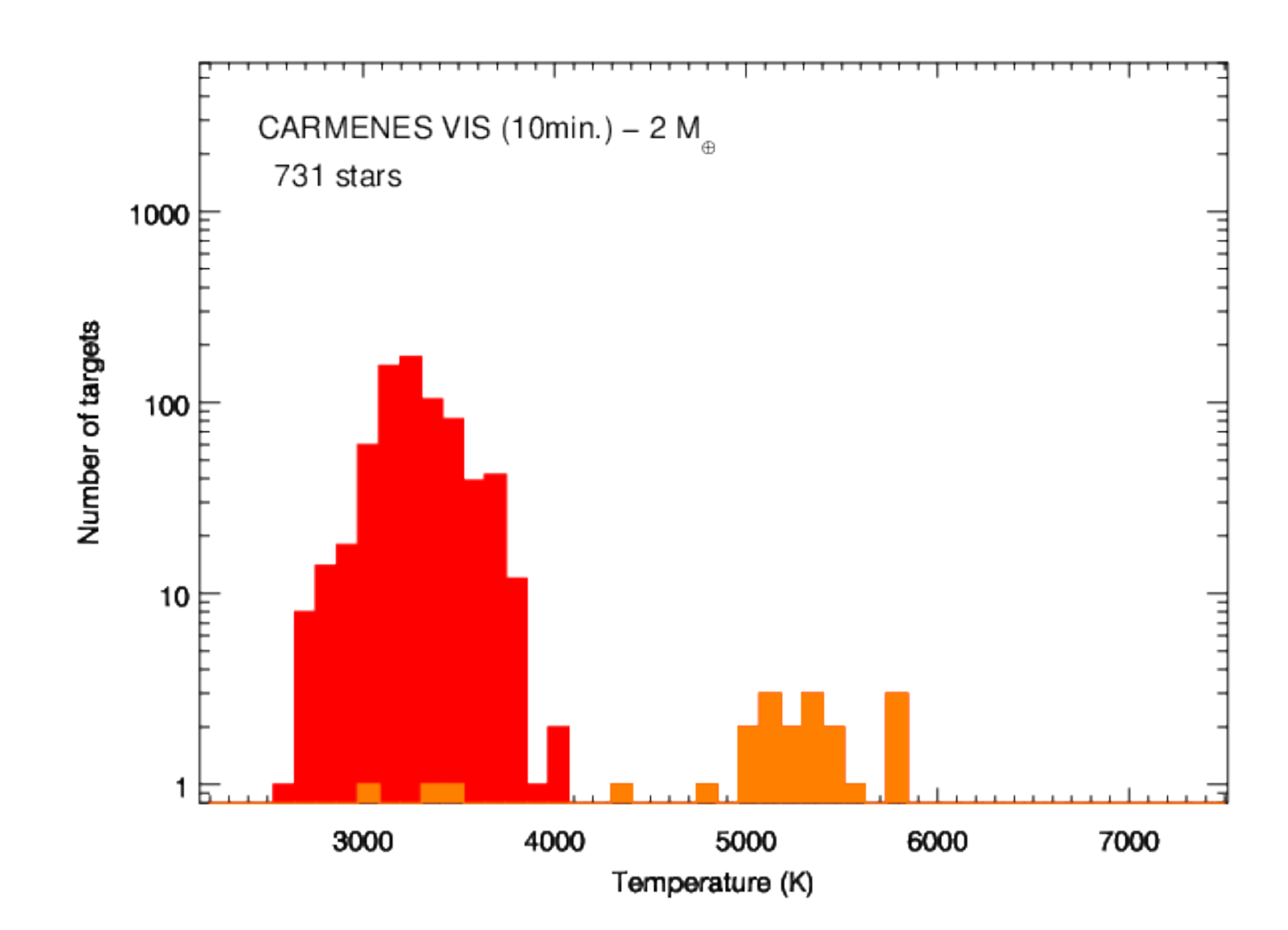}{0.5\textwidth}{(c)}
          \fig{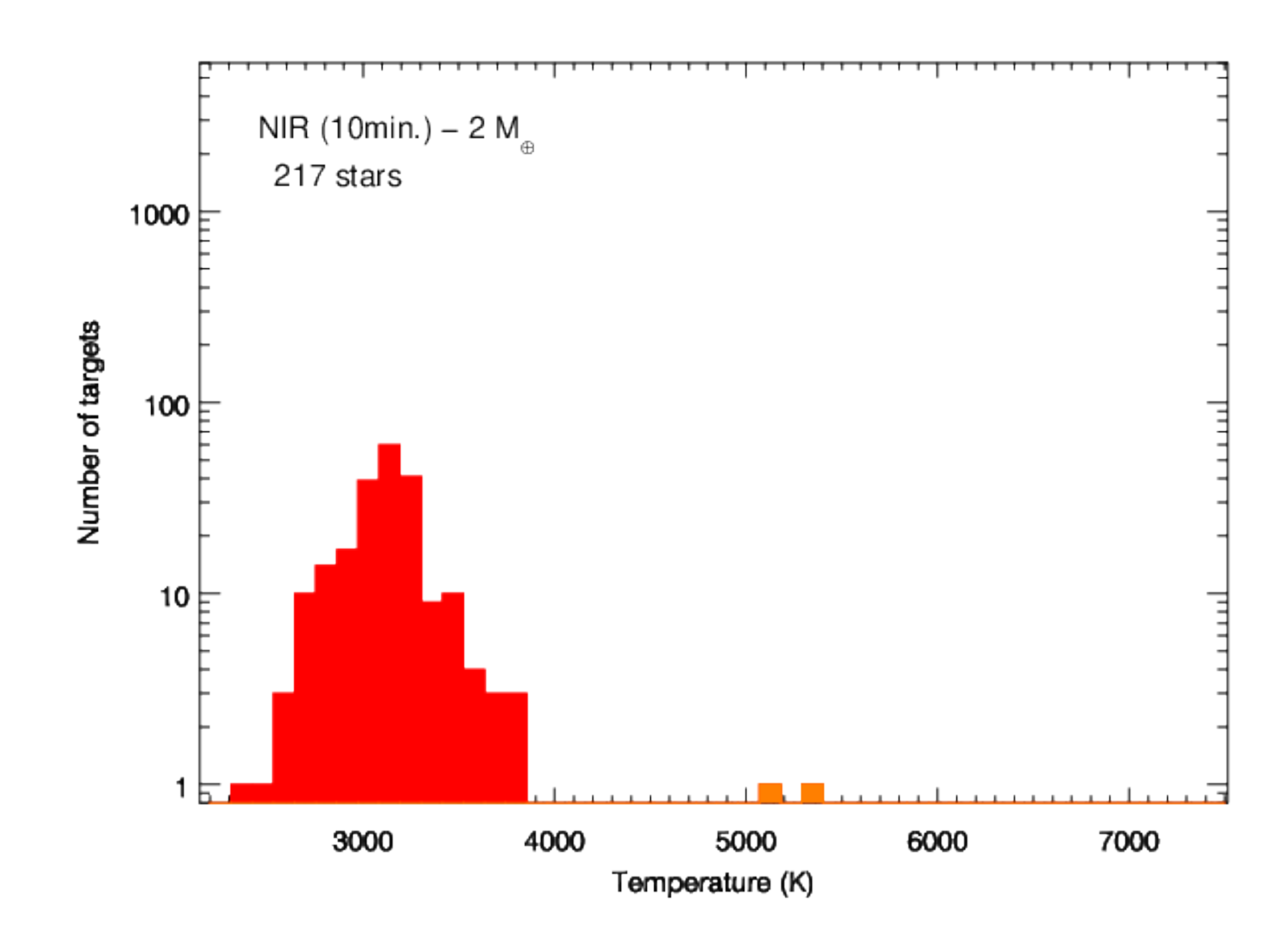}{0.5\textwidth}{(d)}
          }
          \caption{\label{fig:hzhisto}Histogram of stars in which
            planets of masses less or equal 1\,M$_{\Earth}$ (ESPRESSO
            design, upper left panel) and 2\,M$_{\Earth}$ (all other)
            can be discovered inside the HZ of the star in a 5\,min
            (ESPRESSO), 10\,min (HARPS, CARMENES VIS), and NIR
            observation. Colors indicate stars from different
            catalogues as in Fig.\,\ref{fig:HRD}.}
\end{figure*}

We now focus on the search for planets inside the liquid water habitable zone
(HZ) of our sample stars \citep{2014ApJ...787L..29K}. For the distance of the
HZ to the star, we use the runaway greenhouse limit for 0.1\,M$_{\oplus}$,
which relatively well matches the position of Earth in the solar system
\citep[see Fig.\,3 of][]{2014ApJ...787L..29K}. In Fig.\,\ref{fig:hzlimit}, we
show for each star the minimum mass of a planet located inside the HZ that
causes an RV amplitude, $K$, as large as the RV photon limit, $\sigma_{\rm
  RV}$, that can be reached within a 10\,min exposure (5\,min for ESPRESSO),
i.e., RV amplitudes identical to the values as shown in
Fig.\,\ref{fig:rvprecisions}. Stars from Gaia DR2 are not shown here because
we did not attempt a mass estimate for these stars. We realize that our
requirement $K = \sigma_{RV}$ for the determination of a planetary orbit does
not capture the full complexity of RV planet discoveries \citep[see,
e.g.,][]{2017A&A...604A..87G, 2018AJ....155...93C,
  2018AJ....156...82C}. Nevertheless, this simple assumption can serve as a
useful guideline to characterize the sample of stars suitable for RV work with
different instruments.

For the case of ESPRESSO, we see that the RV photon noise potentially
allows the detection of planets inside the HZ that are less massive
than 1\,M$_{\oplus}$ around both sun-like stars and M dwarfs. With the
4m-class instruments, we find that RV photon limits can be reached
that in principle allow to find planets down to 1\,M$_{\oplus}$ and
below around a small sample of M-dwarfs. For sun-like stars, the HARPS
design provides enough information in the brightest targets to see RV
amplitudes caused by planets of a few Earth-masses. The CARMENES VIS
and the NIR designs are less efficient in sun-like stars.

In Fig.\,\ref{fig:hzhisto}, we show histograms of the stars around
which the RV photon noise is as low as the RV amplitude caused by a
planet of $M = 2$\,M$_{\oplus}$ inside the star's HZ if observed with
one of the 4m-class instruments, and for a planet of $M =
1$\,M$_{\oplus}$ for the ESPRESSO design. For the latter, we find 1065
stars that fulfill these criteria including both sun-like stars and
cooler stars. About half the number of stars (493) but with similar
distribution in temperature are available in our sample for HARPS-like
instruments (but for a 2\,M$_{\oplus}$ limit). The distributions of
stars for the two redder designs extend down to lower temperatures but
contain only very few stars hotter than 4000\,K. For the CARMENES VIS
design, we find 731 stars in which low-mass planets inside the HZ
cause an RV amplitude similar to the RV photon limit. For the NIR
design, we find a total number of 217 stars that fulfill our criteria.

\section{Summary}

To answer the questions (a) what are the minimum detectable masses of
planets orbiting the stars in our galactic neighborhood, and (b) with
what kind of spectrograph can they be discovered, we construct a
sample of well characterized dwarf stars with spectral types F--M from
different literature catalogues. This sample contains 46,480 stars and
presumably comprises most of the relevant targets for ongoing and
planned RV programs. We provide an online-tool where RV photon limits
can be calculated for different spectrograph configurations.

Based on empirical information from the HARPS and CARMENES
instruments, together with model simulations, we construct a
consistent set of values for the RV photon noise achievable in
spectroscopic observations at wavelengths between 400\,nm and
2400\,nm. This information is combined with our knowledge about
stellar brightness to estimate the RV photon noise that can be
obtained in an observation with a given instrument and telescope. We
specifically estimate the RV photon noise for four example
configurations: (1) an ESPRESSO-like design at an 8m-class telescope,
(2) an optical spectrograph like HARPS at a 4m-telescope, (3) a red
optical spectrograph like CARMENES-VIS at a 4m-telescope, and (4) a
near-infrared spectrograph covering wavelengths redward of 900\,nm at
a 4m-telescope (similar to CARMENES-NIR, SPIROU, NIRPS, and
GIANO). Comparing instrument performances, we come to the conclusion
that an optical design like HARPS (case 2) provides more RV
information than a near-IR instrument (case 4) in dwarf stars with
temperatures above $\sim$3200\,K. Compared to the red optical
spectrograph (case 3), the optical design (case 2) is more efficient
for stellar temperatures above 4000\,K.

For our catalogue stars, we determine the RV photon noise after 5\,min
observations for the ESPRESSO design, and after 10\,min observations
for the other, and we identify the targets in which this value is
lower than 0.1\,m\,s$^{-1}$ for the ESPRESSO design and 1\,m\,s$^{-1}$
for the other. In addition, we determine the minimum mass of planets
inside the liquid-water habitable zones that would cause RV amplitudes
equal to the RV photon noise for each of the four instrument
configurations. For the ESPRESSO design, we find that more than 280
stars provide enough information in a 5-min observation to push the RV
photon noise below 0.1\,m\,s$^{-1}$. These stars include both M dwarfs
and sun-like stars. For the 4m-class setups (cases 2--4), we find that
the optical, (HARPS) design can reach the RV photon noise limit of
1\,m\,s$^{-1}$ in approx.\ 10,000 stars. For the red-optical
(CARMENES-VIS) design, this number is roughly 3000, and it is about
700 for the NIR design. The optical design is performing best in
sun-like stars but there are also hundreds of targets down to
temperatures of 3000\,K for which this limit is achievable. The
red-optical and near-IR designs can reach low photon limits in even
cooler stars, but all very-low-mass stars at temperatures below
2700\,K are so faint that too few photons for an RV photon limit of
1\,m\,s$^{-1}$ can be collected at any wavelength with a 4m-telescope
in 10\,min exposures.

As for the number of low-mass planets inside the HZ, we find that
those with masses below 1\,M$_{\oplus}$ can cause an RV amplitude
similar to the RV photon noise in more than 1000 stars with the
ESPRESSO-like setup using 5\,min observations. For the optical setup,
we find almost 500 stars in which the RV amplitude caused by a
2\,M$_{\oplus}$ planet inside the HZ would be larger than the RV
photon noise in a 10\,min observation. For both instruments, the
candidate stars include sun-like and low-mass stars. The red-optical
and near-IR setups are optimized for the characterization of low-mass
planets around low-mass stars. For the same criteria (2\,M$_{\oplus}$
planets and 10\,min exposure time), we find more than 700 targets for
the red-optical setup, and more than 200 targets for the near-IR
instruments. With a few exceptions, these stars are M dwarfs cooler
than 4000\,K.

Our estimates did not include the effects of instrument stability,
individual rotation rates, stellar activity, etc., that are all adding
up against the detection of Doppler motion smaller than a few
m\,s$^{-1}$. Improving instrument strategies, long-term stability, and
cross-calibration among different instruments, as well as mitigating
the deteriorating effects of stellar variability are key to the
success of RV missions with the goal of finding very-low mass
planets. Our analysis shows that current technology can obtain enough
spectroscopic information to detect Earth-like planets in a large
number of known stars, and our catalogue should help to identify the
most suitable targets for current and future RV missions and the
search for other Earths.

\acknowledgments

We are very thankful to R.~Cloutier for helpful discussions about
radial velocity noise. The RV precision online calculator was
developed by Daniel Elkeles. We acknowledge financial support from the
DFG Research Unit FOR\,2544 ``Blue Planets around Red Stars''.

%

\bibliography{refs.bib}



\appendix

\section{Star table}


\begin{longrotatetable}
  \begin{deluxetable*}{rCCCCCCCrcccccccc}
    \tablecaption{Catalogue of F--M stars. Targets are taken from the catalogues listed in Table\,\ref{tab:catalogues}. RV limits and HZ-detection limits are for 5\,min in the case of the ESPRESSO design (ES), 10\,min for HARPS (HA) and CARMENES VIS (CV), and 20\,min for the NIR instrument. See Table\,\ref{tab:spectrographs} for a detailed description of the instrument configurations. \label{tab:master}}
    \tabletypesize{\scriptsize}
    \tablewidth{0pt}
    \tablehead{
      \colhead{ID} &
      \colhead{$T_{\rm eff}$} &
      \colhead{d} &
      \colhead{{\rm Mass}} &
      \colhead{$V$} &
      \colhead{$J$} &
      \colhead{$L$} &
      \colhead{HZD} &
      \colhead{Ref} &
      \multicolumn{4}{c}{RV limit (m\,s$^{-1}$)} &
      \multicolumn{4}{c}{HZ-detection limit (M$_{\Earth}$)}\\
      & \colhead{(K)}&
      \colhead{(pc)} & 
      \colhead{(M$_{\odot})$} &
      \colhead{(mag)} &
      \colhead{(mag)} &
      \colhead{(L$_{\odot}$)}&
      \colhead{(AU)}& \ &
      \colhead{ES} & \colhead{HA} & \colhead{CV} & \colhead{NIR} &
      \colhead{ES} & \colhead{HA} & \colhead{CV} & \colhead{NIR}
    }
    \startdata
      2MASS J14294291-6240465 &  2883 &    1.3 &  0.14 & 10.76 &  5.36 &   0.003 &   0.06 & Gai &  0.12 &  0.41 &  0.23 &  0.32 &  0.13 &  0.42 &  0.24 &  0.33 \\
  2MASS J17574849+0441405 &  3237 &    1.8 &  0.14 &  9.49 &  5.24 &   0.003 &   0.06 & Gai &  0.08 &  0.26 &  0.19 &  0.46 &  0.09 &  0.27 &  0.20 &  0.47 \\
  2MASS J10562886+0700527 &  2865 &    2.4 &  0.14 & 12.94 &  7.09 &   0.003 &   0.06 & Gai &  0.32 &  1.14 &  0.56 &  0.72 &  0.33 &  1.17 &  0.58 &  0.74 \\
  2MASS J11032023+3558117 &  3593 &    2.5 &  0.46 &  7.51 &  4.20 &   0.028 &   0.18 & Gai &  0.05 &  0.17 &  0.13 &  0.31 &  0.17 &  0.54 &  0.42 &  1.00 \\
  2MASS J18494929-2350101 &  3213 &    3.0 &  0.14 & 10.41 &  6.22 &   0.003 &   0.06 & Gai &  0.14 &  0.44 &  0.31 &  0.73 &  0.14 &  0.45 &  0.32 &  0.75 \\
  2MASS J23415498+4410407 &  3005 &    3.2 &  0.14 & 12.41 &  6.88 &   0.003 &   0.06 & Gai &  0.23 &  0.79 &  0.46 &  0.75 &  0.24 &  0.82 &  0.48 &  0.77 \\
                 HD 22049 &  5116 &    3.2 &  0.82 &  3.72 &       &   0.388 &   0.65 & Nor &  0.03 &  0.05 &  0.11 &  0.19 &  0.21 &  0.44 &  0.92 &  1.55 \\
  2MASS J11474440+0048164 &  3145 &    3.4 &  0.14 & 11.21 &  6.51 &   0.003 &   0.06 & Gai &  0.17 &  0.54 &  0.37 &  0.76 &  0.17 &  0.56 &  0.38 &  0.78 \\
                 HD 61421 &  6683 &    3.5 &  1.53 &  0.37 &       &   3.934 &   1.89 & Nor &  0.08 &  0.16 &  0.43 &  0.44 &  1.50 &  3.11 &  8.11 &  8.41 \\
  2MASS J18424688+5937374 &  3334 &    3.5 &  0.25 & 10.00 &  5.72 &   0.009 &   0.10 & Gai &  0.10 &  0.33 &  0.25 &  0.58 &  0.19 &  0.60 &  0.45 &  1.04 \\
  2MASS J00182256+4401222 &  3669 &    3.6 &  0.50 &  8.15 &  5.25 &   0.036 &   0.21 & Gai &  0.09 &  0.29 &  0.23 &  0.53 &  0.32 &  1.03 &  0.82 &  1.91 \\
  2MASS J00182549+4401376 &  3282 &    3.6 &  0.19 & 11.08 &  6.79 &   0.006 &   0.09 & Gai &  0.18 &  0.58 &  0.42 &  0.96 &  0.25 &  0.83 &  0.59 &  1.36 \\
  2MASS J08294949+2646348 &  2809 &    3.6 &  0.14 & 14.16 &  8.23 &   0.003 &   0.06 & Gai &  0.69 &  2.72 &  1.09 &  1.24 &  0.71 &  2.80 &  1.12 &  1.28 \\
  2MASS J18424666+5937499 &  3392 &    3.6 &  0.30 &  8.92 &  5.19 &   0.013 &   0.13 & Gai &  0.08 &  0.25 &  0.19 &  0.45 &  0.17 &  0.55 &  0.42 &  0.98 \\
                 HD 10700 &  5345 &    3.6 &  0.81 &  3.50 &       &   0.451 &   0.69 & Nor &  0.03 &  0.05 &  0.12 &  0.20 &  0.22 &  0.46 &  0.97 &  1.64 \\
  2MASS J01123052-1659570 &  3062 &    3.7 &  0.14 & 11.96 &  7.26 &   0.003 &   0.06 & Gai &  0.27 &  0.92 &  0.55 &  0.97 &  0.27 &  0.94 &  0.56 &  1.00 \\
  2MASS J03355969-4430453 &  2999 &    3.7 &  0.14 & 13.06 &  7.52 &   0.003 &   0.06 & Gai &  0.33 &  1.17 &  0.64 &  1.01 &  0.34 &  1.20 &  0.66 &  1.04 \\
  2MASS J07272450+0513329 &  3317 &    3.8 &  0.23 &  9.81 &  5.71 &   0.008 &   0.10 & Gai &  0.10 &  0.33 &  0.25 &  0.57 &  0.17 &  0.55 &  0.41 &  0.96 \\
  2MASS J02530084+1652532 &  2700 &    3.9 &  0.14 & 14.51 &  8.39 &   0.003 &   0.06 & Gai &  1.06 &  4.67 &  1.59 &  1.25 &  1.09 &  4.81 &  1.64 &  1.28 \\
  2MASS J05114046-4501051 &  3695 &    3.9 &  0.52 &  8.93 &  5.82 &   0.039 &   0.22 & Gai &  0.12 &  0.39 &  0.30 &  0.71 &  0.44 &  1.45 &  1.14 &  2.66 \\
  2MASS J21171534-3852022 &  3776 &    3.9 &  0.56 &  6.75 &  4.05 &   0.048 &   0.24 & Gai &  0.05 &  0.16 &  0.14 &  0.31 &  0.22 &  0.66 &  0.57 &  1.27 \\

    \enddata
  \end{deluxetable*}
\end{longrotatetable}

\end{document}